\documentclass[twocolumn]{aastex6}

\let\pwiflocal=\iffalse \let\pwifjournal=\iffalse
\usepackage{amsmath,amssymb}
\usepackage{epsfig}    
\usepackage{graphicx}    
\usepackage{lineno}
\usepackage{natbib}
\usepackage{bigints}
\usepackage[outdir=./]{epstopdf}

\usepackage[T1]{fontenc}
\pwifjournal\else
  \usepackage{microtype}
\fi

\pwifjournal\else
  \makeatletter
  \renewcommand\plotone[1]{%
    \centering \leavevmode \setlength{\plot@width}{0.95\linewidth}
    \includegraphics[width={\eps@scaling\plot@width}]{#1}%
  }%
  \makeatother
\fi

\makeatletter

\newcommand\@simpfx{http://simbad.u-strasbg.fr/simbad/sim-id?Ident=}

\newcommand\MakeObj[4][\@empty]{
  \pwifjournal%
    \expandafter\newcommand\csname pkgwobj@c@#2\endcsname[1]{\protect\object[#4]{##1}}%
  \else%
    \expandafter\newcommand\csname pkgwobj@c@#2\endcsname[1]{\href{\@simpfx #3}{##1}}%
  \fi%
  \expandafter\newcommand\csname pkgwobj@f#2\endcsname{#4}%
  \ifx\@empty#1%
    \expandafter\newcommand\csname pkgwobj@s#2\endcsname{#4}%
  \else%
    \expandafter\newcommand\csname pkgwobj@s#2\endcsname{#1}%
  \fi}%

\newcommand\MakeTrunc[2]{
  \expandafter\newcommand\csname pkgwobj@t#1\endcsname{#2}}%

\newcommand{\obj}[1]{%
  \expandafter\ifx\csname pkgwobj@c@#1\endcsname\relax%
    \textbf{[unknown object!]}%
  \else%
    \csname pkgwobj@c@#1\endcsname{\csname pkgwobj@s#1\endcsname}%
  \fi}
\newcommand{\objf}[1]{%
  \expandafter\ifx\csname pkgwobj@c@#1\endcsname\relax%
    \textbf{[unknown object!]}%
  \else%
    \csname pkgwobj@c@#1\endcsname{\csname pkgwobj@f#1\endcsname}%
  \fi}
\newcommand{\objt}[1]{%
  \expandafter\ifx\csname pkgwobj@c@#1\endcsname\relax%
    \textbf{[unknown object!]}%
  \else%
    \csname pkgwobj@c@#1\endcsname{\csname pkgwobj@t#1\endcsname}%
  \fi}

\makeatother

\pwifjournal\else
  \usepackage{etoolbox}
  \makeatletter
  \patchcmd{\NAT@citex}
    {\@citea\NAT@hyper@{%
       \NAT@nmfmt{\NAT@nm}%
       \hyper@natlinkbreak{\NAT@aysep\NAT@spacechar}{\@citeb\@extra@b@citeb}%
       \NAT@date}}
    {\@citea\NAT@nmfmt{\NAT@nm}%
     \NAT@aysep\NAT@spacechar\NAT@hyper@{\NAT@date}}{}{}
  \patchcmd{\NAT@citex}
    {\@citea\NAT@hyper@{%
       \NAT@nmfmt{\NAT@nm}%
       \hyper@natlinkbreak{\NAT@spacechar\NAT@@open\if*#1*\else#1\NAT@spacechar\fi}%
         {\@citeb\@extra@b@citeb}%
       \NAT@date}}
    {\@citea\NAT@nmfmt{\NAT@nm}%
     \NAT@spacechar\NAT@@open\if*#1*\else#1\NAT@spacechar\fi\NAT@hyper@{\NAT@date}}
    {}{}
  \makeatother
\fi

\providecommand{\adsurl}[1]{\href{#1}{ADS}}

\newcommand{\vsini}{$v\sin{i_*}$}

\newcommand{\kepler}{{\it Kepler}}

\newcommand{\um}{$\mu$m}
\newcommand{\fbol}{$F_{\mathrm{bol}}$}

\newcommand\teff{\ensuremath{T_\text{eff}}}

\newcommand\kms{km~s$^{-1}$}

\newcommand{\ktwo}{{\it K2}}
\newcommand{\Kepler}{{\it Kepler}}

\newcommand{\tar}{K2-136}
\newcommand{\ms}{\,m\,s$^{-1}$}
\newcommand{\teq}{T$_\textrm{eq}$}

\newcommand\ut{1}
\newcommand\hub{3}
\newcommand\col{2}
\newcommand\har{4}

\newcommand\sagan{5}

\slugcomment{Submitted to Journals of the AAS}

\shorttitle{Three-planet system in Hyades}

\shortauthors{Mann et al.}

\begin{document}
 
\title{Zodiacal Exoplanets in Time (ZEIT) VI:\\ a three-planet system in the Hyades cluster including an Earth-sized planet} 

\author{Andrew W. Mann\footnote{\email{awm2126@columbia.edu}}\altaffilmark{\ut,\col,\hub}, 
Andrew Vanderburg\altaffilmark{\ut,\har,\sagan}, 
Aaron C. Rizzuto\altaffilmark{\ut}, 
Adam L. Kraus\altaffilmark{\ut}, 
Perry Berlind\altaffilmark{\har}, 
Allyson Bieryla\altaffilmark{\har}, 
Michael L. Calkins\altaffilmark{\har}, 
Gilbert A. Esquerdo\altaffilmark{\har}, 
David W. Latham\altaffilmark{\har}, 
Gregory N. Mace\altaffilmark{\ut}, 
Nathan R. Morris\altaffilmark{\ut}, 
Samuel N. Quinn\altaffilmark{\har}, 
Kimberly R. Sokal\altaffilmark{\ut}, 
Robert P. Stefanik\altaffilmark{\har} 
}

\altaffiltext{\ut}{Department of Astronomy, The University of Texas at Austin, Austin, TX 78712, USA}
\altaffiltext{\col}{Department of Astronomy, Columbia University, 550 West 120th Street, New York, NY 10027, USA}
\altaffiltext{\hub}{NASA Hubble Fellow}
\altaffiltext{\har}{Harvard--Smithsonian Center for Astrophysics, Cambridge, Massachusetts 02138, USA}
\altaffiltext{\sagan}{NASA Sagan Fellow}

\begin{abstract}
Planets in young clusters are powerful probes of the evolution of planetary systems. Here we report the discovery of three planets transiting \tar\ (EPIC 247589423), a late K dwarf in the Hyades ($\simeq$800\,Myr) cluster, and robust detection limits for additional planets in the system. The planets were identified from their {\it K2} light curves as part of our survey of young clusters and star forming regions. The smallest planet has a radius comparable to Earth ($0.99^{+0.06}_{-0.04}R_\oplus$), making it one of the few Earth-sized planets with a known, young age. The two larger planets are likely a mini-Neptune and a super-Earth, with radii of $2.91^{+0.11}_{-0.10}R_\oplus$ and $1.45^{+0.11}_{-0.08}R_\oplus$, respectively. The predicted radial velocity signals from these planets are between 0.4 and 2 m/s, achievable with modern precision RV spectrographs. Because the target star is bright ($V$=11.2) and has relatively low-amplitude stellar variability for a young star (2-6\,mmag), \tar\ hosts the best known planets in a young open cluster for precise radial velocity follow-up, enabling a robust test of earlier claims that young planets are less dense than their older counterparts.
\end{abstract}



\maketitle

\section{Introduction}\label{sec:intro}
NASA's {\it Kepler} mission \citep{Borucki2010} has massively expanded our understanding of the final configuration of planetary systems, in large part by enabling population studies based on large datasets. {\it Kepler} results, accompanied by a wide range of ground-based follow-up, have facilitated studies of small-planet occurrence \citep[e.g.,][]{Fressin:2013qy,Muirhead2015}, detailed correlations between stellar and planet properties \citep[e.g.,][]{Mann2013b,2017ApJ...838...25G}, and the mass-radius relation for small exoplanets \citep[e.g.,][]{2014ApJ...783L...6W,2015ApJ...801...41R}, among a wide range of other planetary and stellar topics. The {\it Kepler} dataset is likely to remain critical for statistics for the foreseeable future, thanks in part to a sensitive and public planet-search pipeline \citep{Jenkins:2010qy,2010ApJ...713L..87J}\footnote{\href{https://github.com/nasa/kepler-pipeline}{https://github.com/nasa/kepler-pipeline}}, detailed analysis of the pipeline completeness \citep{2016ApJ...828...99C}, and the upcoming arrival of {\it Gaia} parallaxes \citep{2012Ap&SS.341...31D}, which is expected to solve many earlier complications assigning stellar parameters to {\it Kepler} target stars \citep[e.g.,][]{Gaidos2013,Bastien2014,Newton2015A}.

After the failure of two reaction wheels, the repurposed {\it Kepler} \citep[{\it K2},][]{Howell2014} has built on {\it Kepler}'s success at exoplanet discovery, primarily by observing populations of stars that were missed or poorly sampled by the original {\it Kepler} mission. {\it K2} is surveying stars that are statistically brighter than the {\it Kepler} mission \citep{Crossfield:2016aa,2014ApJS..211....2H,Huber2016}, enabling more detailed follow-up with ground based resources. Compared to {\it Kepler}, {\it K2} targets also include a much larger sample of late-type stars \citep{Dressing2017}, white dwarfs \citep{2015Natur.526..546V,2017MNRAS.468.1946H}, and disk-bearing stars \citep{2016ApJ...816...69A}.

Of particular importance for studies of exoplanet evolution, {\it K2} has also surveyed a number of young ($<1$\,Gyr) clusters and star forming regions, facilitating the discovery of planets across a wide range of ages. While the {\it Kepler} mission did include open clusters, these are older and more distant \citep{2011ApJ...733L...9M,Meibom2013}, and hence less useful for studies of the critical early stages of an exoplanet's life. {\it K2}, however, observed parts of Hyades and Praesepe \citep[800\,Myr,][]{Brandt2015}, Pleiades \citep[112\,Myr,][]{2015ApJ...813..108D}, Upper Scorpius \citep[11\,Myr,][]{2012ApJ...746..154P}, and Taurus-Auriga \citep[0-5\,Myr,][]{2017ApJ...838..150K}. Numerous earlier radial velocity (RV) studies have found planets in these regions \citep[e.g.,][]{Quinn2014,2016A&A...588A.118M,Donati:2017aa}, however, the \ktwo\ sample now represents the majority of the young cluster planets, as well as the only non-Jovian planets, in these regions \citep[e.g.,][]{David2016b,Gaidos:2017aa,Mann:2017aa}, and has enabled stellar and exoplanetary science beyond what is possible with RV surveys \citep[e.g.,][]{2016ApJ...822...47D,2017arXiv170603084G,2017ApJ...845...72K}

Of the young planets discovered to date, three (K2-25b, K2-33b, and K2-95b) have radii significantly larger than planets orbiting stars of similar mass ($M_*<0.6M_\odot$) at comparable orbital periods \citep[e.g.,][]{Obermeier:2016aa,2017AJ....153..177P}. K2-100 also has an unusually high radius given the level of incident flux from its host star \citep{2016NatCo...711201L,2017AJ....153..271S}. Together this suggests that planets are less dense when young, however, the origin and prevalence of this difference in radii is unclear. \citet{Mann2016a} argue for ongoing atmospheric escape due to interactions with the active host star, as has been seen for some older planets \citep{Ehrenreich2015}. However, the amount of the mass-loss required complicates this explanation; for example, K2-25b is $\simeq4R_\earth$, while all close-in planets orbiting similar-mass stars discovered by {\it Kepler} are $\le$2$R_\earth$ \citep{Dressing2013,Gaidos2016b}. Given that K2-25 is $\simeq$800\,Myr old, this suggests mass-loss rates larger than expected from the high-energy flux from the host \citep{2012ApJ...761...59L} and beyond what has been observed around even relatively active stars \citep[e.g.,][]{2010ApJ...717.1291L,2011A&A...529A.136E}.

Mass determinations for some of the young cluster planets would help to answer questions about both the occurrence and physical driver of the observed radius differences. However, with the exception of Hyades, all the young clusters and star forming regions probed by {\it K2} are more than 100\,pc from the Sun, and a disproportionate fraction of the known young planets orbit fainter M dwarfs. This puts nearly all appropriate targets outside the range of existing RV spectrographs ($V\lesssim12$). Only K2-100b \citep{Mann:2017aa} orbits a star bright enough for a precise RV mass in the immediate future, which still would leave masses for the bulk of the young planetary sample unknown. 

Here we present the discovery of a three-planet system orbiting \tar\ (EPIC 247589423, LP 358-348), a late-K dwarf in the Hyades star cluster. The smallest and shortest-period planet is Earth-sized, making it the first Earth-sized planet found in a young cluster, while the two outer planets are a super-Earth and a mini-Neptune. Because these planets orbit a bright (V$\simeq$11) star that is relatively quiet for its age, \tar\ currently represents the best cluster target for dedicated RV observations to measure the planetary masses. The host star's small size and proximity also make atmospheric characterization possible with {\it HST} and {\it JWST}. 

The paper is structured as follows. We present our reduction of the {\it K2} light curve, method for identifying the transits, and ground-based follow-up observations in Section~\ref{sec:obs}. In Section~\ref{sec:params} we derive parameters of the host star, including membership to the Hyades cluster. Using an injection/recovery test, we test our sensitivity to additional planets around \tar\ in Section~\ref{sec:limit}. We detail our transit-fitting procedure in Section~\ref{sec:transit}, and assess the probability that the transit signals are false positives in Section~\ref{sec:false}. We summarize and discuss our findings in Section~\ref{sec:discussion}, including a brief discussion of the prospects for measuring the masses (and therefore densities) of these new planets. 

\section{Observations and Data Reduction}\label{sec:obs} 

\subsection{\ktwo\ Observations and Transit Identification}\label{sec:k2}
\ktwo\ observed \tar\ from 8 March 2017 to 27 May 2017 during Campaign 13. The raw pixel-level data were calibrated by the \Kepler\ pipeline \citep{2010SPIE.7740E..1UT,Stumpe2012}\footnote{\href{https://keplerscience.arc.nasa.gov/pipeline.html}{https://keplerscience.arc.nasa.gov/pipeline.html}} and released publicly on 28 August 2017. \tar\ was selected by seven different guest observer programs including our own (GO13008, GO13018, GO13023, GO13049, GO13064, GO13077, and GO13090)\footnote{\href{https://keplerscience.arc.nasa.gov/k2-approved-programs.html}{https://keplerscience.arc.nasa.gov/k2-approved-programs.html}} six of which selected \tar\ because of its known membership to Hyades \citep{Roser2011,2013A&A...559A..43G}. 


\begin{figure*}
\begin{center}
\includegraphics[width=0.9\textwidth]{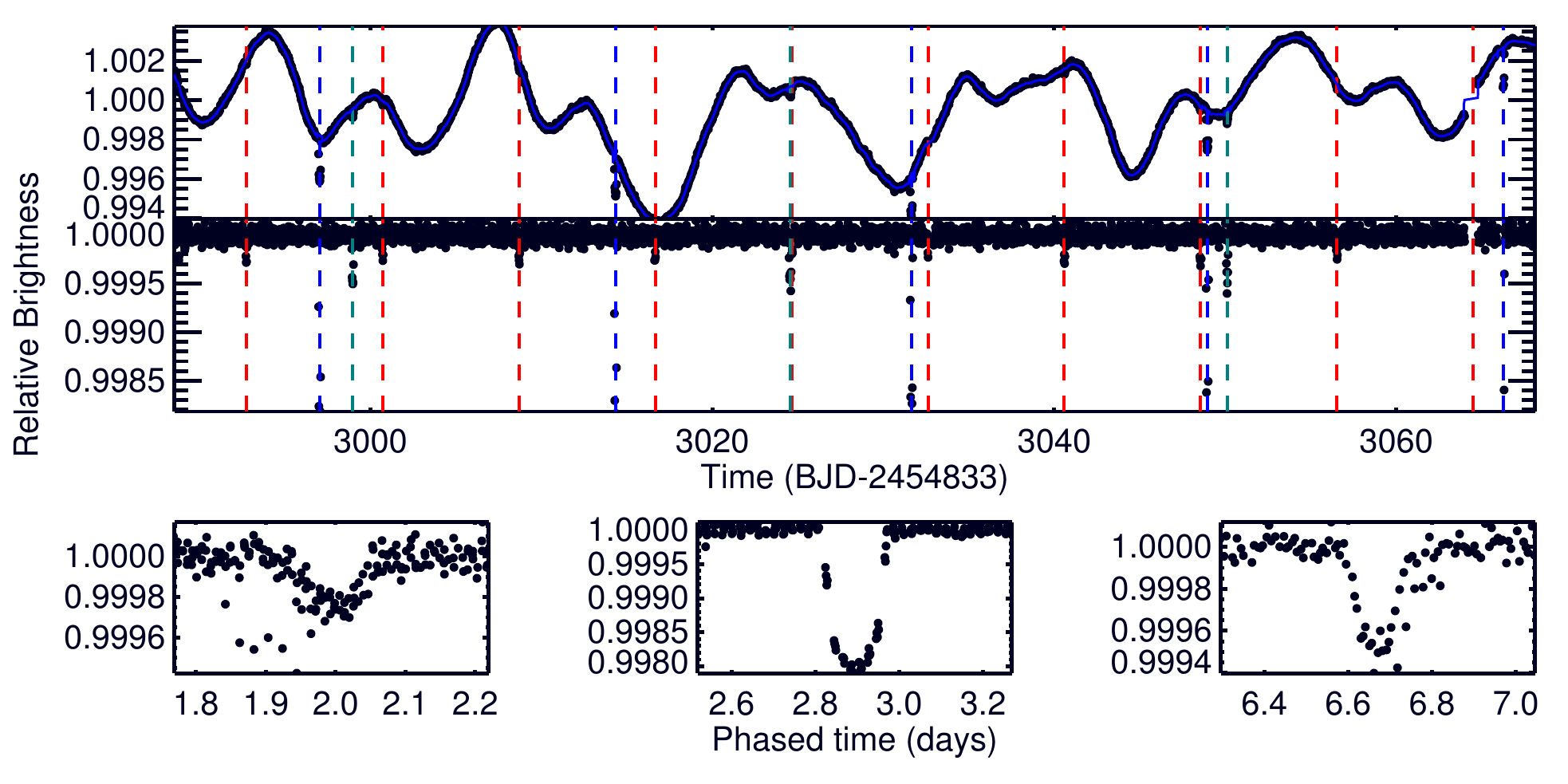}
\caption{Light curve of \tar\ from {\it K2}. The top panel shows the light curve after correcting for the {\it K2} roll/flat field. The center panel shows the same curve after correcting for stellar variability with the planets masked (post-identification). Dashed lines mark transits, color-coded by the planet. The bottom three panels show the final phase-folded transits of each planet in order of period (shortest period on the left). Outliers (primarily flares) have been removed from the phase-folded light curves. Low points present in the b and d phased light curves just after and before the transit are due to partially overlapping transits. }
\label{fig:lc}
\end{center}
\end{figure*}

The details of our \ktwo\ transit-search pipeline are described in detail in \citet{2017arXiv170909670R}, including our methods for removing young star variability. To briefly summarize, we first corrected for systematic errors caused by \ktwo's unstable pointing and pixel-response variations following the method of \citet{Vanderburg2014}. Once we produced a light curve mostly free of instrumental systematics, we removed astrophysical stellar variability using a `notch' filter, which fits small regions of the light curve separately using the combination of an outlier-resistant polynomial and a trapezoidal-shaped notch. Including the notch prevented our otherwise aggressive smoothing algorithm from removing transits, and mitigated skewing of the stellar variability fit due to the presence of a transit. After the identification of each planet, in-transit points were masked out and an additional search is performed. In this way we identified three signals around \tar, at periods of $\simeq$17.3 days, $\simeq25.6$ days, and $\simeq$7.9 days (in order of significance). We inspected each signal by eye and determined them to be consistent with planetary signals, and hence worthy of further investigation. We show the light curve and detected signals in Figure~\ref{fig:lc}. 

After we identified the transiting planets, we re-extracted the light curve using a simultaneous least-squares fit to the transits, low-frequency variability, and K2 roll systematics, coupled with outlier removal (primarily flares), as described by \citet{Becker2015} and \citet{v16}. The final light curve (flattened by removing the best-fit low-frequency variability) was used for our MCMC transit-fit, as described in Section~\ref{sec:transit}.

\subsection{Archival Spectroscopy with the CfA Digital Speedometers}

\tar\ was observed five times as part of a RV survey of Hyades cluster members between November 1983 and January 2004 with the CfA Digital Speedometer spectrograph on the 1.5m Tillinghast reflector at Fred L. Whipple Observatory on Mt. Hopkins, AZ (three times) and on the 1.5m Wyeth reflector at Oak Ridge Observatory in the town of Harvard, Massachusetts (twice). The observations showed no significant RV variations at the level of roughly $\simeq$0.5\,\kms. The average absolute RV of \tar\ as measured by the CfA Digital Speedometers was 39.7\,\kms, consistent with membership in the Hyades cluster.  

\subsection{Optical Spectra from TRES}\label{sec:tres}

We observed \tar\ with the Tillinghast Reflector Echelle Spectrograph on the 1.5m telescope at Fred L. Whipple Observatory twice: on 2 September 2017 and 4 September 2017. The second observation had a signal-to-noise ratio of about 28 per resolution element at 520 nm, while the first observation was taken in mediocre conditions and had a lower signal-to-noise ratio of about 16 per resolution element. Both spectra are available on the Exoplanet Follow-up Observing Program for \ktwo\ website\footnote{\href{https://exofop.ipac.caltech.edu/k2/edit_target.php?id=247589423}{https://exofop.ipac.caltech.edu/k2/}}.

RVs were determined from the TRES spectra using the Stellar Parameter Classification tool \citep[SPC, ][]{2012Natur.486..375B,2014Natur.509..593B}. SPC measured absolute RVs from the spectra by cross-correlating with a library of synthetic spectra generated from \citet{kurucz1992} stellar atmosphere models (varying effective temperature, velocity, and line broadening) and selecting the template which gave the strongest cross-correlation peak. After applying a correction to place the TRES velocities on the IAU absolute frame, this yielded a RV of 39.73 and 39.76 \kms\ for the first and second observation, respectively. Both values are consistent with the measurements from the Digital Speedometers. The uncertainty on these absolute velocities is 100-150\ms, primarily due to uncertainties in the IAU absolute velocity scale.

\subsection{High-resolution Spectra with IGRINS}\label{sec:IGRINS}
We observed \tar\ on the night of 2017 Sep 1 (UT) with the Immersion Grating Infrared Spectrometer \citep[IGRINS,][]{2010SPIE.7735E..1MY,Park2014} on the Discovery Channel Telescope located at Happy Jack, AZ. IGRINS provides high resolving power ($R\simeq$45,000) and simultaneous coverage of both $H$ and $K$ bands (1.48-2.48\,\um) with RV stability of $\lesssim$40\ms\ using telluric lines for wavelength calibration \citep{Mace2016}.

All observations were taken following commonly used strategies for point-source observations with IGRINS, including observing along two positions on the slit to facilitate sky subtraction, and observing an A0V telluric standard immediately before the target at a similar airmass \citep{Vacca2003}. The final spectrum has a SNR of $>$200 (per resolution element) in the peak of the $H$- and $K$-bands. The spectrum was reduced using version 2.2 of the publicly available IGRINS pipeline package\footnote{https://github.com/igrins/plp} \citep{IGRINS_plp}. 

Barycentric RVs were derived from the IGRINS spectra following the method outlined in \citet{Mann2016a}. To briefly summarize, we used the telluric spectrum to correct the wavelength solution, then cross-correlated each of 42 orders against a grid of RV templates of similar spectral type to \tar, all taken with IGRINS. The final RV and error is the mean and standard deviation of the values from all 44 templates, with a zero-point term (and associated error) to place all RVs on the same scale, yielding a final RV of 39.10$\pm$0.20\,\kms. This is slightly lower but consistent with the values from other sources and the expected velocity for Hyades cluster membership.

\subsection{Literature Photometry and Astrometry}

We compiled optical $BVg'r'i'$ photometry from the eighth data release of the AAVSO All-Sky Photometric Survey \citep[APASS;][]{Henden:2012fk}, NIR $JHK_S$ photometry from The Two Micron All Sky Survey \citep[2MASS;][]{Skrutskie2006}, $r'$ photometry from the 15th Carlsberg Meridian Catalog \citep[CMC15,][]{CMC15}, {\it Gaia} $G$ from the first {\it Gaia} data release \citep[DR1, ][]{gaiadr1}, and $W1-4$ infrared photometry from the Wide-field Infrared Survey Explorer \citep[WISE;][]{Wright2010}. The WISE $W4$ magnitude is an upper limit only (non-detection), and was discarded. The APASS magnitudes were measured from a single epoch; errors are estimated only from measurement (mostly Poisson) errors ($\simeq$0.01 for each band). Because of stellar and atmospheric variability, these errors are likely underestimated. We instead adopted errors of 0.08\,mags based on a comparison of APASS single-epoch magnitudes to those from the Sloan Digital Sky Survey \citep[SDSS,][]{Ahn:2012kx}. 

We drew proper motions from ``Hot Stuff For One Year'' \citep[HOSY,][]{hsoy}, which combined proper motions/astrometry from {\it Gaia} DR1 and PPMXL \citep{2010AJ....139.2440R}, which itself was built by joining USNO-B1.0 \citep{Monet:2003fj} and 2MASS. HSOY reports a proper motion of 83.0$\pm$0.9, -35.7$\pm$0.9\,mas/yr for \tar, consistent with values from other proper motion sources \citep[e.g.,][]{Roser2011,2017AJ....153..166Z,URAT1}. 
 
These basic data on \tar\ are given in Table~\ref{tab:stellar}, with all RVs in Table~\ref{tab:rv}.

\section{Stellar Parameters}\label{sec:params} 

\begin{figure*}[htbp]
\begin{center}
\includegraphics[width=0.46\textwidth]{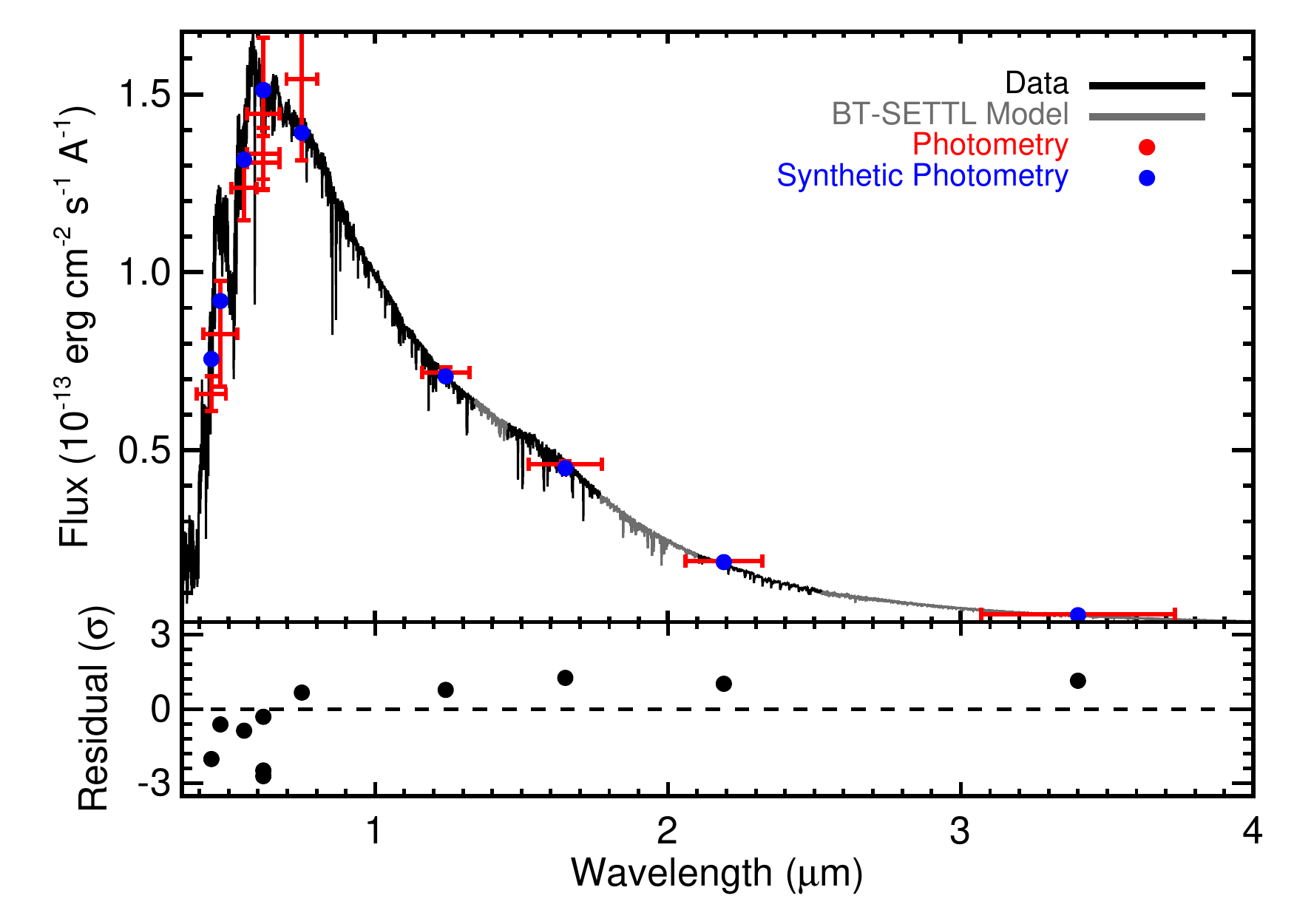}
\includegraphics[width=0.46\textwidth]{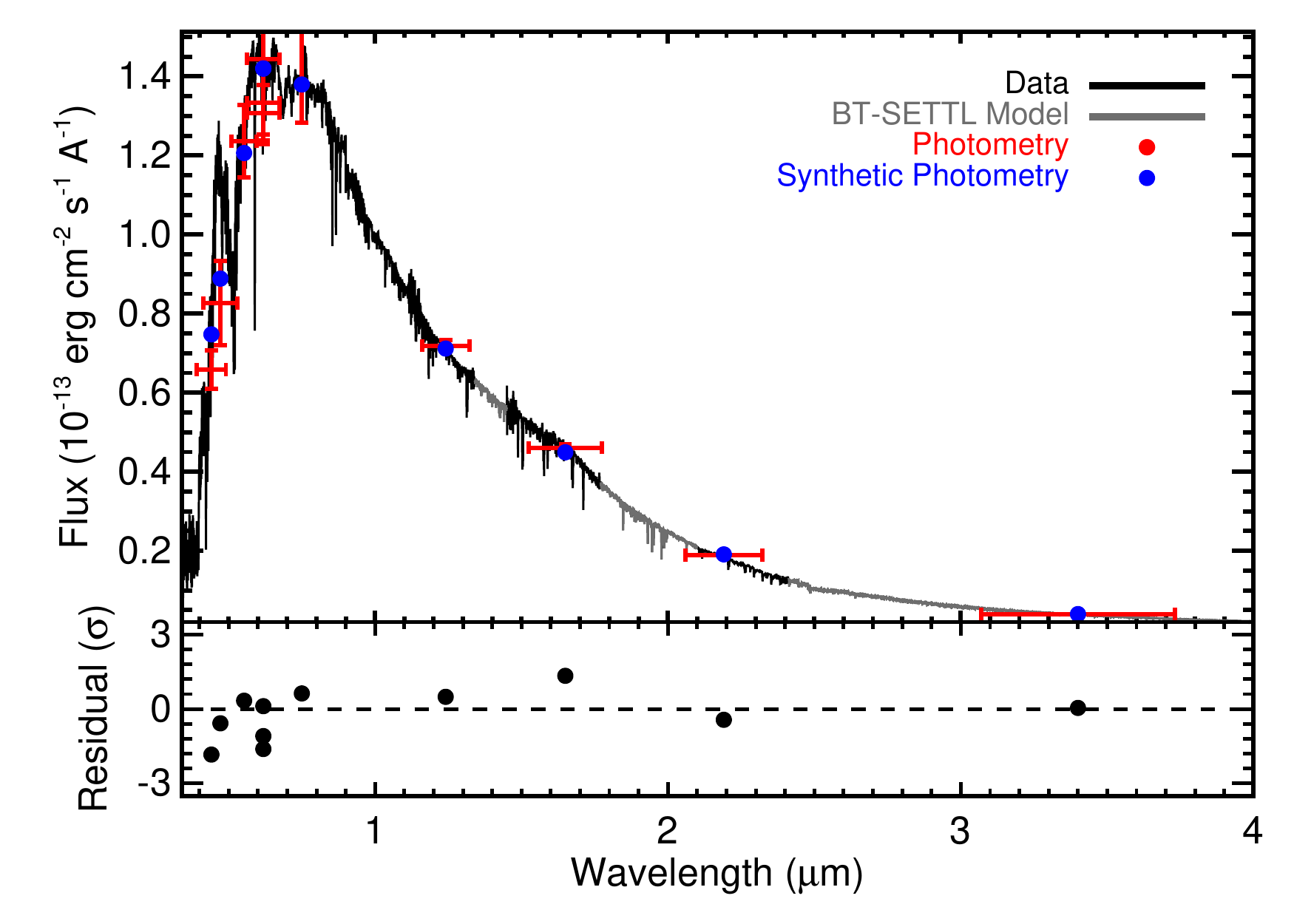}
\caption{Template optical and NIR spectra (black) fit to the archival photometry. We filled in regions lacking spectra or with high telluric contamination with models (grey). Literature photometry is shown as red points, with errors. The X-axis error bars approximate the effective width of the filter. Blue points are the synthetic photometry from the template spectrum. The left shows the best-fit K5 template, with the best-fit K7 template is on the right. }
\label{fig:sed}
\end{center}
\end{figure*}

{\it Spectral Type and Bolometric Flux:} Our technique for measuring the bolometric flux (\fbol) is described in significant detail in \citet{Mann2015b}. The main difference here is that we did not have flux-calibrated spectra of \tar, and instead used a grid of template spectra to fit the archival photometry. We pulled NIR spectral templates from \citet{Rayner2009}, with complementary optical spectra taken from \citet{Lepine:2013} or \citet{2014MNRAS.437.3133G}. We converted spectra to magnitudes using the appropriate filter profiles and zero points with corresponding errors \citep{2003AJ....126.1090C,Mann2015a}. BT-SETTL models \citep{Allard2011} were used to fill gaps in our template spectra, including regions of poor atmospheric transmission. We also allowed small (1-3\%) corrections to the flux calibration in regions of overlap between the optical and NIR spectra as part of the fit. We repeated this process for each template from K3 to M3, each time comparing the synthetic (computed from templates) magnitudes to the archive photometry. We achieved reasonable fits with K5 (reduced $\chi^2 = 2.6$) and K7 (reduced $\chi^2=1.1$) templates (see Figure~\ref{fig:sed}), and poor fits with all others (reduced $\chi^2>6$). For our final spectral type we adopted K5.5$\pm$0.5.

As part of our fit, each template was shifted to match the flux levels of the observed photometry (absolute flux calibration). During this process, we computed \fbol\ by integrating over the full template and atmospheric model. We use the $\chi^2$-weighted \fbol\ between the K5 and K7 templates as our final value, yielding a bolometric flux of  0.149$\pm0.005\times10^{-8}$~erg\,cm$^{-2}$\,s$^{-1}$.

\begin{figure}
\begin{center}
\includegraphics[width=0.46\textwidth]{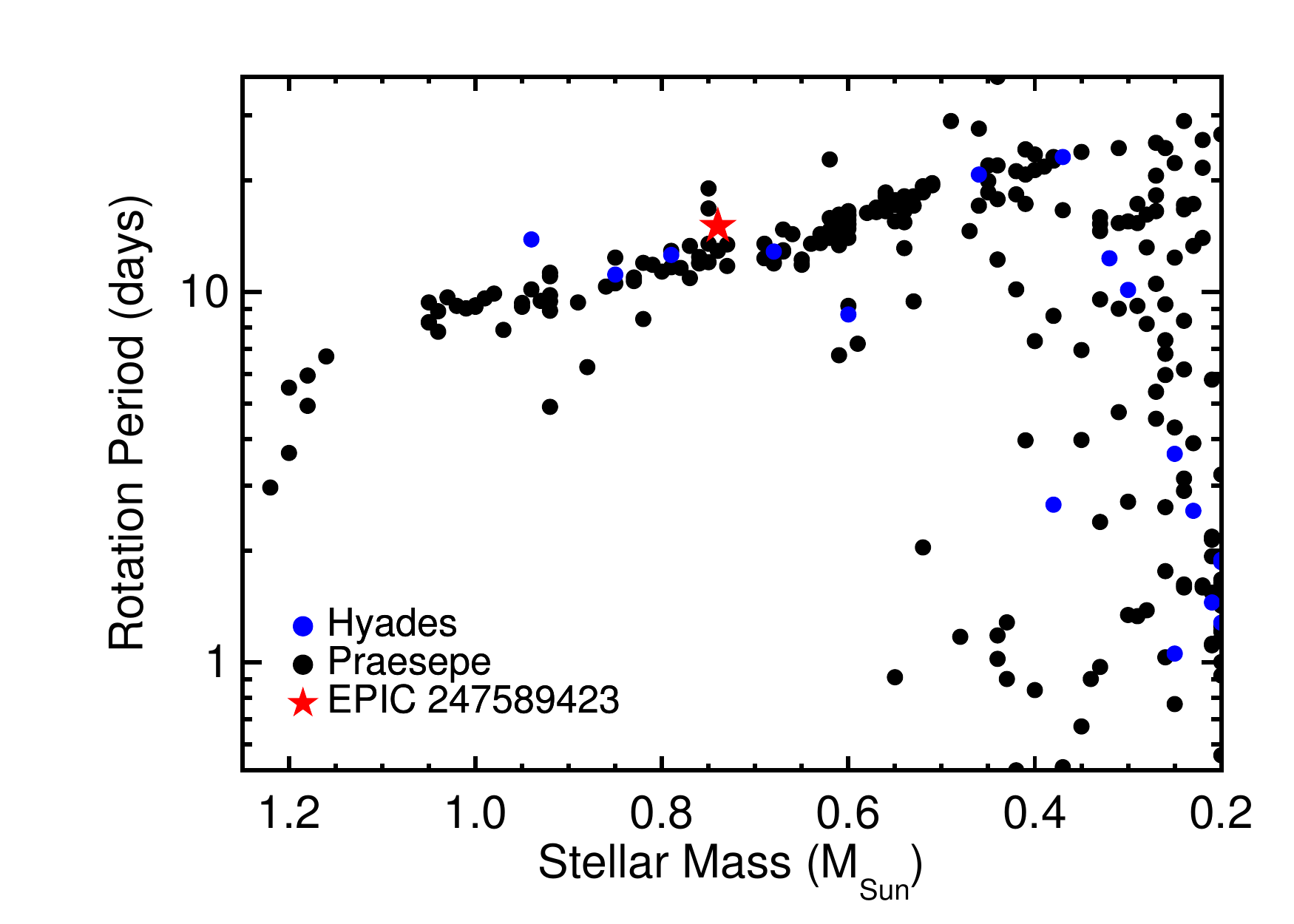}
\caption{Rotation period versus mass of \tar\ (red star) compared to that of other stars in the Hyades (blue) and Praesepe (red) clusters. Rotation periods for stars other than \tar\ were taken from \citet{2016ApJ...822...47D} and \citet{2017ApJ...842...83D} after excluding known binaries.}
\label{fig:rot}
\end{center}
\end{figure}

{\it Rotation period}: We computed a Lomb-Scargle periodogram for periods of 1-30\,days from the {\it K2} light curve (prior to applying stellar variability correction). We fit the largest peak in the power spectrum with a Gaussian, adopting the center and width as the rotation period and a conservative estimate of the error (15.04$\pm$1.01\,days). The second highest peak lands at $\simeq7.5$ days, corresponding to a harmonic of the true rotation period. As a check, we also measured the rotation period using the SuperSmoother code \citep{jake_vanderplas_2015_14475}, which gave a consistent value. Our adopted rotation period sits on the high end of the sequence compared to similar-mass members of Hyades and Praesepe clusters \citep{2017ApJ...845...72K,2016ApJ...822...47D}, although not significantly so (Figure~\ref{fig:rot}).

{\it TRES model-fitting:} Our SPC analysis of the TRES spectrum is described in Section~\ref{sec:tres}, which was used to derive RVs. We performed a second SPC analysis to determine spectroscopic parameters. Here, SPC cross-correlated the TRES spectra with synthetic template spectra varying temperature, gravity, metallicity, and line broadening and then interpolated the height of the peaks as a function of these parameters to find the best values. The output parameters were \teff=4499$\pm$50\,K, log~g=4.64$\pm$0.10, [m/H]=-0.10$\pm$0.08, and \vsini=3.0$\pm$0.5\,km/s. The template spectra used by SPC include a 1.9 \kms\ microturbulence term, but no macroturbulence. Because both turbulence terms are not known precisely in young stars, our errors on \vsini\ may be underestimated somewhat.

\begin{figure}[htbp]
\begin{center}
\includegraphics[width=0.46\textwidth]{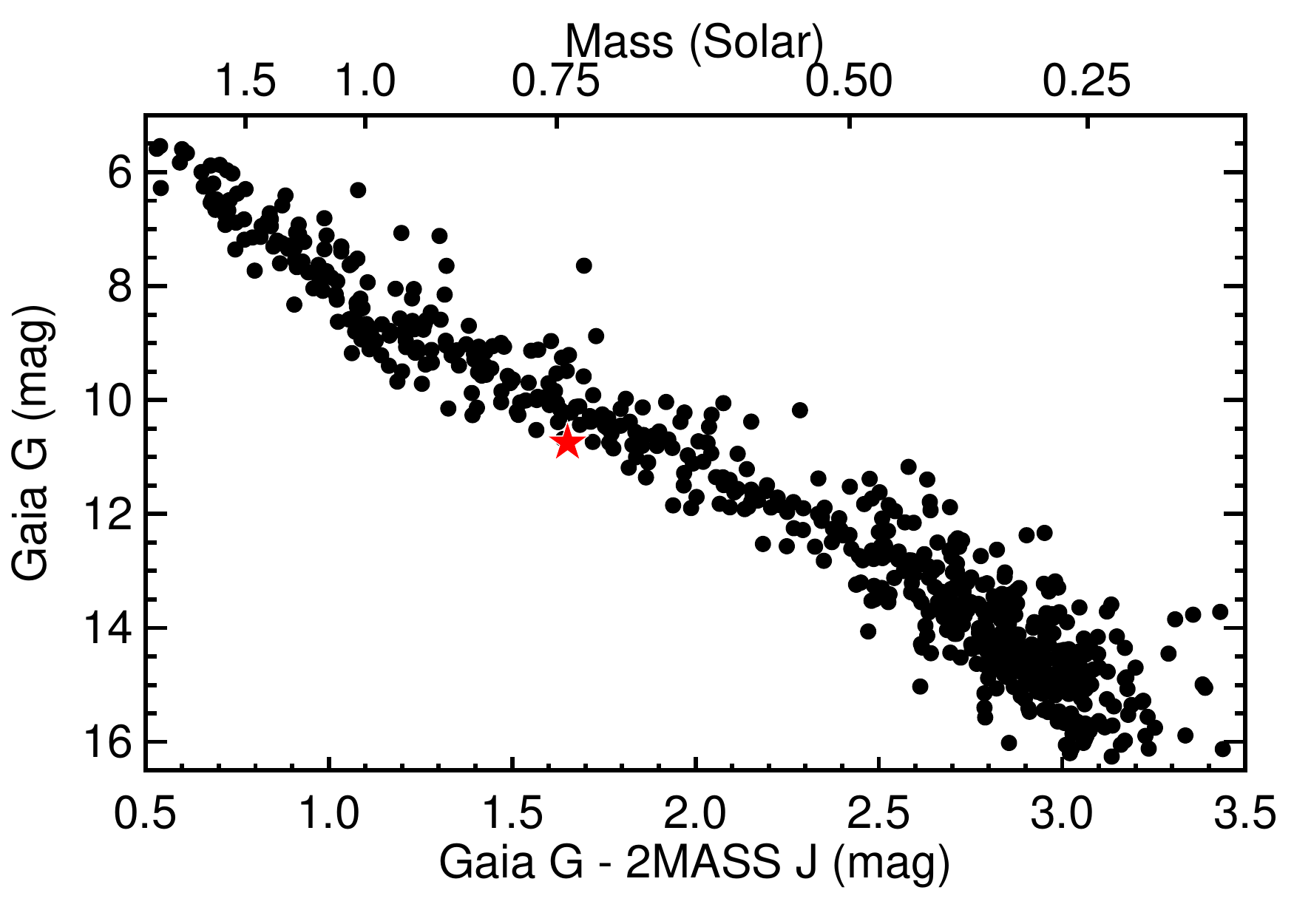}
\caption{Color-magnitude diagram of Hyades cluster members with $G$ magnitudes from $Gaia$ and $J$ from 2MASS. The red star indicates the location of \tar. Approximate masses are shown on the top axis. Typical errors on {\it Gaia} magnitudes are $<0.01$, and $\simeq0.02$ for 2MASS $J$, generally smaller than the width of the points. Most of the scatter is due to spread in individual cluster member distances and binarity. A tight, high-density line of points can be seen corresponding to single-stars within the cluster core; \tar\ falls below (more distant than) this core, consistent with its 3-dimensional position (Figure~\ref{fig:xyz}).  }
\label{fig:cmd}
\end{center}
\end{figure}

{\it Membership to the Hyades cluster:} The kinematics, position, and photometry of \tar\ are highly consistent with membership in the Hyades open cluster. In the absence of a measured parallax for \tar, we calculated a photometric distance of 63$\pm$10\,pc from the (G-J,G) \citep{gaiadr1,Skrutskie2006} CMD of the Hyades cluster (Figure~\ref{fig:cmd}). This is consistent with the distances to the Hyades population ($\sim$47\,pc, \citealt{Roser2011}). At this photometric distance, the proper motions taken from HSOY \citep{hsoy} ($\mu_\alpha,\mu_\delta=83.0\pm0.9$, $-35.7\pm0.9$\,mas/yr), agree with the expected cluster proper motions projected from the Hyades Galactic space velocity (E($\mu_\alpha,\mu_\delta$)=($81\pm9$, $-30\pm9$)\,mas/yr). Similarly, the projected cluster RV of 37.3$\pm$2\,km/s is consistent with the RV of 39.7$\pm$0.1\,km/s measured for \tar. Using the Bayesian method from \citet{rizzuto2011,rizzuto2015}, and the Hyades cluster model from \citet{2017arXiv170909670R}, we calculated a Hyades cluster membership probability of $>$99\% for \tar.

\begin{figure*}[htbp]
\begin{center}
\includegraphics[width=0.97\textwidth]{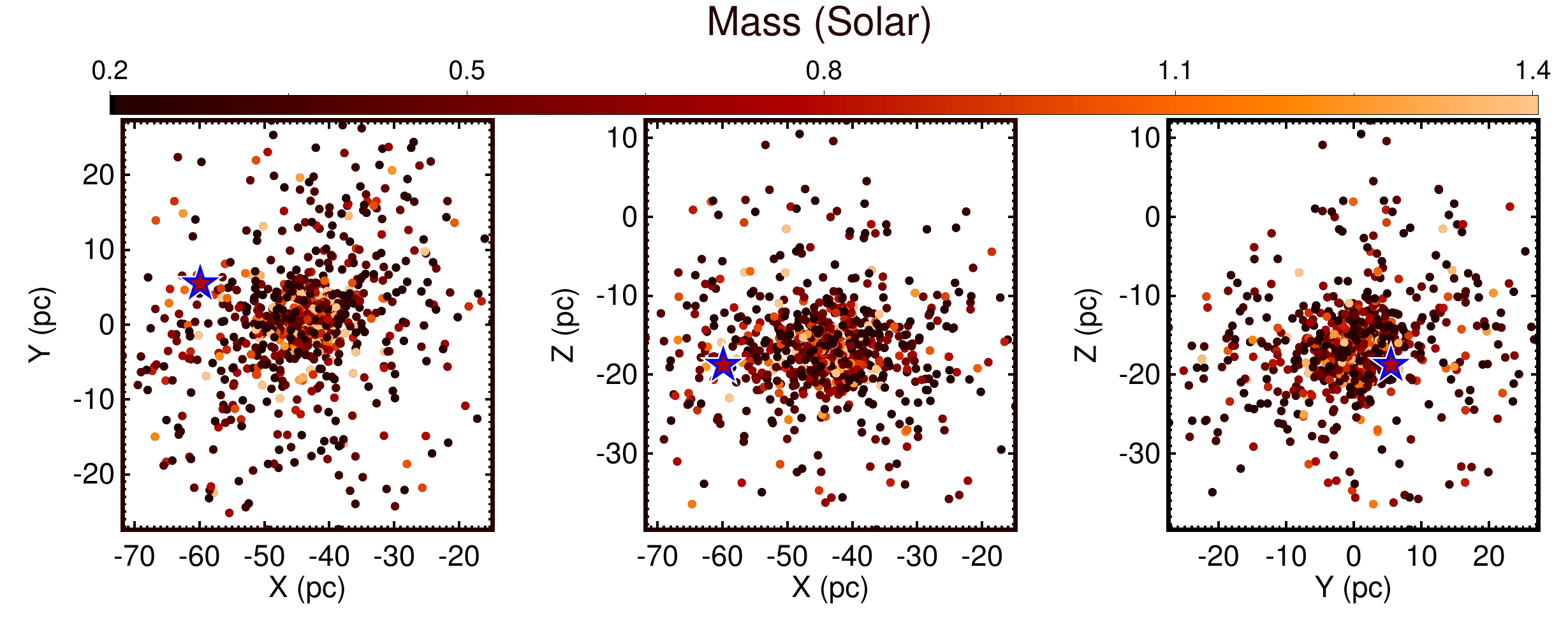}
\caption{Galactic coordinates ($XYZ$) of Hyades members colored by their model-estimated mass. \tar\ is shown as a star with a white/blue outline. \tar\ is within the greater Hyades cluster population, but outside the denser cluster core, consistent with its CMD position.} 
\label{fig:xyz}
\end{center}
\end{figure*}

Combined with a rotation period consistent with the Hyades sequence (Figure~\ref{fig:rot}), we conclude that \tar\ is a member of Hyades, in agreement with previous determinations \citep{Roser2011,2013A&A...559A..43G}. Based on \tar's Galactic position ($XYZ$), the star lands 10-20\,pc from the cluster center and outside of the $\simeq$5\,pc radius core (Figure~\ref{fig:xyz}), but still well within the broader population of Hyades members. 

{\it Metallicity:} We adopted the Hyades cluster metallicity for \tar. For the cluster value, we used the literature determinations \citep{2003AJ....125.3185P,Brandt2015,Dutra-Ferreira2016}, based primarily on spectroscopic measurements of higher-mass stars. The range of these literature values is 0.12~$<$~[Fe/H]~$<$~0.18, with typical uncertainties of 0.03~dex. We used [Fe/H]=0.15$\pm$0.03, which encompasses these measurements and errors that may be common to all determinations. The [m/H] value from our TRES spectrum is below this ([m/H]=-0.10$\pm$0.10) at the $\gtrsim2\sigma$ level (taking [Fe/H]$\simeq$[m/H]), but we consider the overall cluster measurements to be more accurate and precise than that of an individual star, and variations in abundance patterns across cluster members have been shown to be smaller than or comparable to our adopted errors \citep{Liu2016}. Further, metallicity determinations for these late-type dwarfs are complicated \citep[e.g.,][]{Neves2012,Mann2013a}, and the models used with SPC do not contain the full suite of molecular bands that begin to appear at the effective temperature of \tar.

{\it Kinematic distance:} Although \tar\ does not have a parallax, membership to the Hyades cluster enables us to calculate the target's distance, from which we can interpolate precise stellar parameters from the luminosity/absolute magnitudes. The distance to the Hyades cluster core is precisely known \citep{van-Leeuwen2009,gaiadr1}, however, Hyades members have been found $>15$\,pc away from the core ($47.5$\,pc from the Sun), suggest errors of $\gtrsim$30\% if applied to individual stars, and \tar\ does not appear to be in the cluster core (Figure~\ref{fig:xyz}). However, a kinematic distance, i.e., the distance that yields Galactic kinematics ($UVW$) consistent with the cluster or moving group \citep[e.g.,][]{Roser2011,Malo2013}, can yield distances to individual stars with uncertainties below 10\%. 

To calculate the kinematic distance to \tar, we used the established Galactic kinematics of Hyades from \citet{van-Leeuwen2009}, allowing for a variation of 1.2\,km/s in the cluster velocity due to dispersion from internal kinematics \citet{2014A&A...564A..49P}. Accounting for errors in the proper motion and RV of \tar, we derive a final distance of $d = 59.4 \pm 2.8$\,pc. 

{\it Distance-based parameters:} We first combined our kinematic distance with our \fbol\ determination to calculate a total stellar luminosity of $L_{bol} = 0.163 \pm 0.016 L_\odot$. 

To calculate $R_*$, $M_*$, and \teff, we interpolated absolute $J$, $H$, and $K_S$ magnitudes onto the 800\,Myr Solar-metallicity isochrones from \citet{BHAC15}. We used NIR $JHK_S$ magnitudes because they are relatively insensitive to reddening, have precise zero-points and filter profiles \citep{2003AJ....126.1090C,Mann2015a}, and yield stellar parameters in good agreement with each other and with the luminosity derived using our \fbol. Switching to the 625\,Myr or 1\,Gyr isochrones did not change any inferred parameter significantly, nor did using using the MESA isochrones \citep[MIST, ][]{2016ApJ...823..102C,2016ApJS..222....8D}. Stellar density ($\rho_*$) and log~g were derived using the $M_*$ and $R_*$, accounting for correlated errors between these values (as they both depend directly on the parallax). 

{\it Comparison of parameters:} Our final adopted \teff\ and log~g values are in excellent agreement with the values derived using the TRES spectra; 4547$\pm83$~K and 4.66$\pm$0.02 from the isochrones versus 4499$\pm$50\,K and 4.64$\pm$0.10 from model fitting to TRES spectra. Further, combining the TRES \teff\ with our \fbol\ and kinematic distance yield a radius of 0.67$\pm0.03R_\odot$ (via Stefan-Boltzmann), in excellent agreement with our isochronal value (0.66$\pm0.02R_\odot$). The consistency between these values from methods adds credence to our small error bars, which are relatively small given the complexity of measuring stellar parameters of late-type dwarfs.

For parameters with multiple determinations, we adopted the more precise value for all analysis. For \teff\ and \vsini, we used values from TRES spectra, but for all other parameters we used those from the kinematic distance. 

A summary of final derived stellar parameters and errors is given in Table~\ref{tab:stellar}.

\section{Limits on additional planets around \tar}\label{sec:limit}

\begin{figure}
\begin{center}
\includegraphics[width=0.45\textwidth,trim={0 150 0 150},clip]{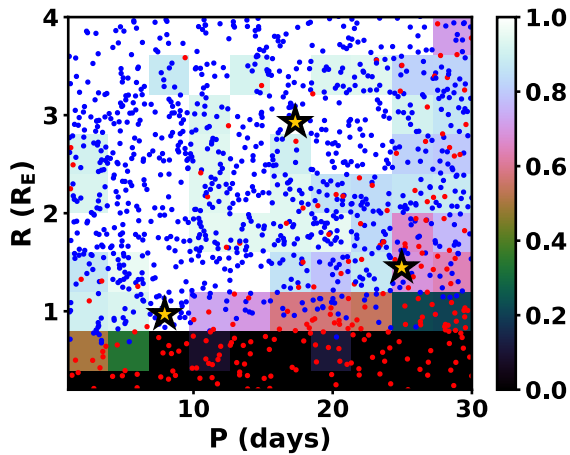}
\caption{Sensitivity of our pipeline to additional planets around \tar, based on injection/recovery tests performed as described by \citet{2017arXiv170909670R}. Blue points correspond recovered planets, while red points are those we missed. Color-coding corresponds to the fraction of planets recovered within a given bin. The orange/black stars mark the locations of the real planets, \tar bcd.}
\label{fig:completeness}
\end{center}
\end{figure}

Following \citep{2017arXiv170909670R}, we set limits on the size and period of any undetected planets around \tar. To this end, we performed a full injection/recovery test on the light curve. We first inserted fake planets into the uncorrected light curve (extracted pixel photometry). We then applied the corrections for {\it K2} pointing and stellar variability and masked out the planets identified above. Using a box-least-squares algorithm \citep{Kovacs2002} on the corrected light curve, we search for planetary signals with SNR$>$7. If a planet was detected at the injected period and phase we consider this recovered.

We repeated this process 2000 times, each time adjusting the period and transit depth. Injected planet parameters were randomly selected, but forced to sample the parameter space. Significantly more information on our injection/recovery test can be found in \citet{2017arXiv170909670R}.

Measurement of our sensitivity based on the injection/recovery test is shown Figure~\ref{fig:completeness}. We find that in general, we are sensitive to sub-Earth-sized planets in short periods ($\lesssim$10\,days), Earth-sized planets in orbiting \tar\ in periods out to $\approx$20 days, and planets $\simeq2R_\oplus$ or larger out to the limit of our search (30 days). This is significantly higher sensitivity than we achieved on earlier ZEIT planet-hosts \citep{2017arXiv170909670R}, thanks to the host star's brightness, small stellar radius, and low stellar variability. 

\section{Transit Fitting}\label{sec:transit}
We fit the \ktwo\ light curve with a Markov Chain Monte Carlo (MCMC) as described in \citet{Mann2016a} and expanded in \citet{Mann:2017ab} for multi-planet systems. To briefly summarize, our fitting code is based on the combination of model light curves produced by the \textit{batman} package \citep{Kreidberg2015} with the {\it emcee} affine invariant MCMC \citep{Foreman-Mackey2013}. Long integration times can distort the light curve \citep{Kipping:2010lr}, so we over-sample and bin the model to match the 30\,min cadence of \ktwo. We assumed a quadratic limb-darkening law with the sampling method of \citet{Kipping2013}. The free parameters for each planet are planet-to-star radius ratio ($R_P/R_*$), impact parameter ($b$), orbital period ($P$), epoch of the first transit mid-point ($T_0$), and two parameters that describe the eccentricity and argument of periastron ($\sqrt{e}\sin\omega$ and $\sqrt{e}\cos\omega$). All planets are fit simultaneously, assuming a common bulk stellar density ($\rho_*$) and two (quadratic) limb-darkening parameters ($q1$ and $q2$).  

We applied Gaussian priors on limb-darkening and stellar density, but all other parameters evolved under uniform priors with only physical boundaries (e.g., $0<R_P/R_*<1$, $-1<b<1$). Limb-darkening priors were derived from the \citet{2013A&A...553A...6H} atmospheric models using the LDTK code \citep{2015MNRAS.453.3821P}, the \kepler\ filter response, and the adopted stellar parameters from Section~\ref{sec:params} and account for errors in stellar parameters and differences in model grids. The applied Gaussian prior was $0.68\pm0.05$ for $u_1$ and $0.09\pm0.05$ for $u_2$. These values and errors are converted to the triangular sampling parameters $q1$ and $q2$ \citep{Kipping2013}. For stellar density we apply a Gaussian prior using our values from Section~\ref{sec:params}.

\begin{figure*}[t]
\begin{center}
\includegraphics[width=0.46\textwidth]{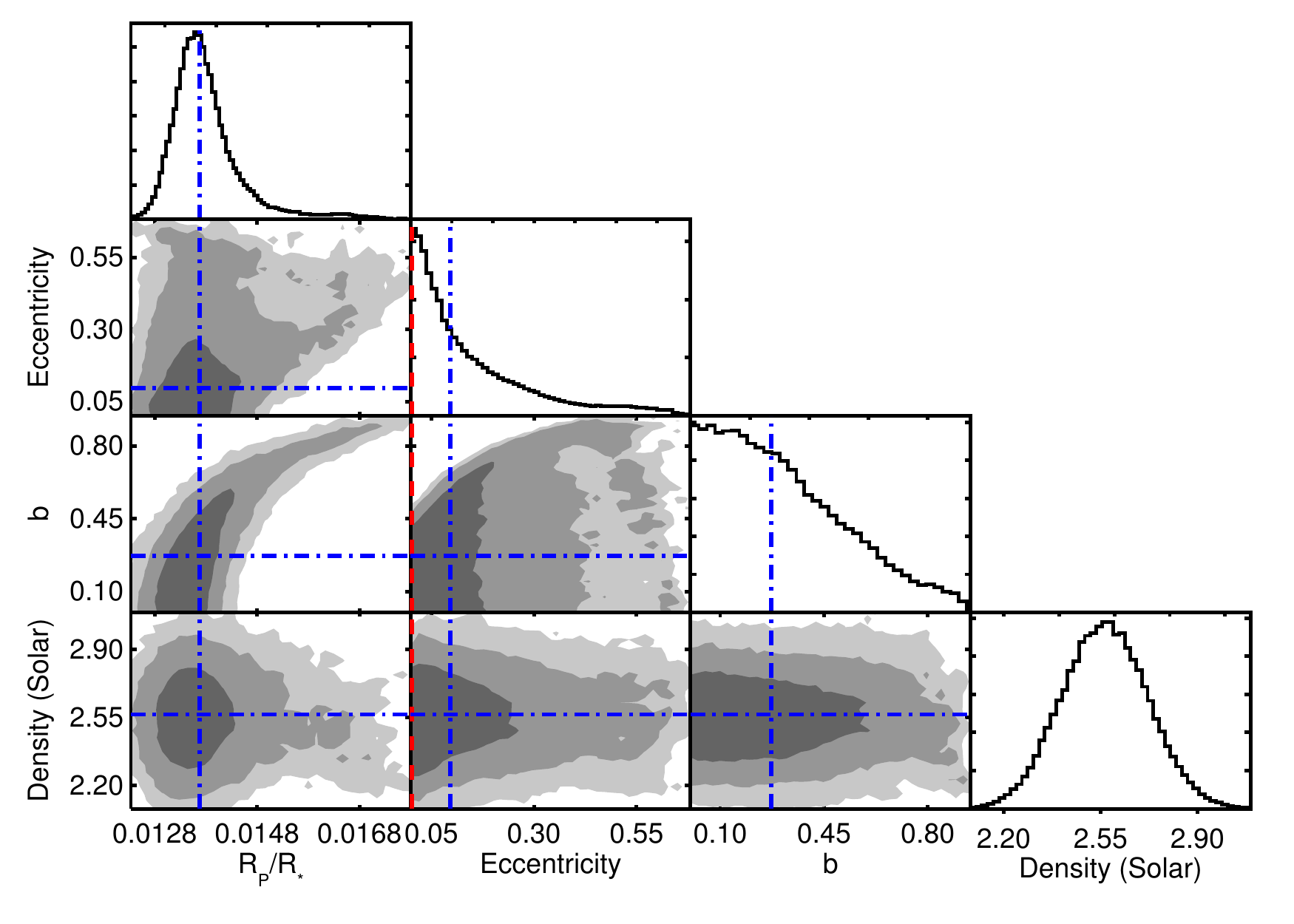}
\includegraphics[width=0.46\textwidth]{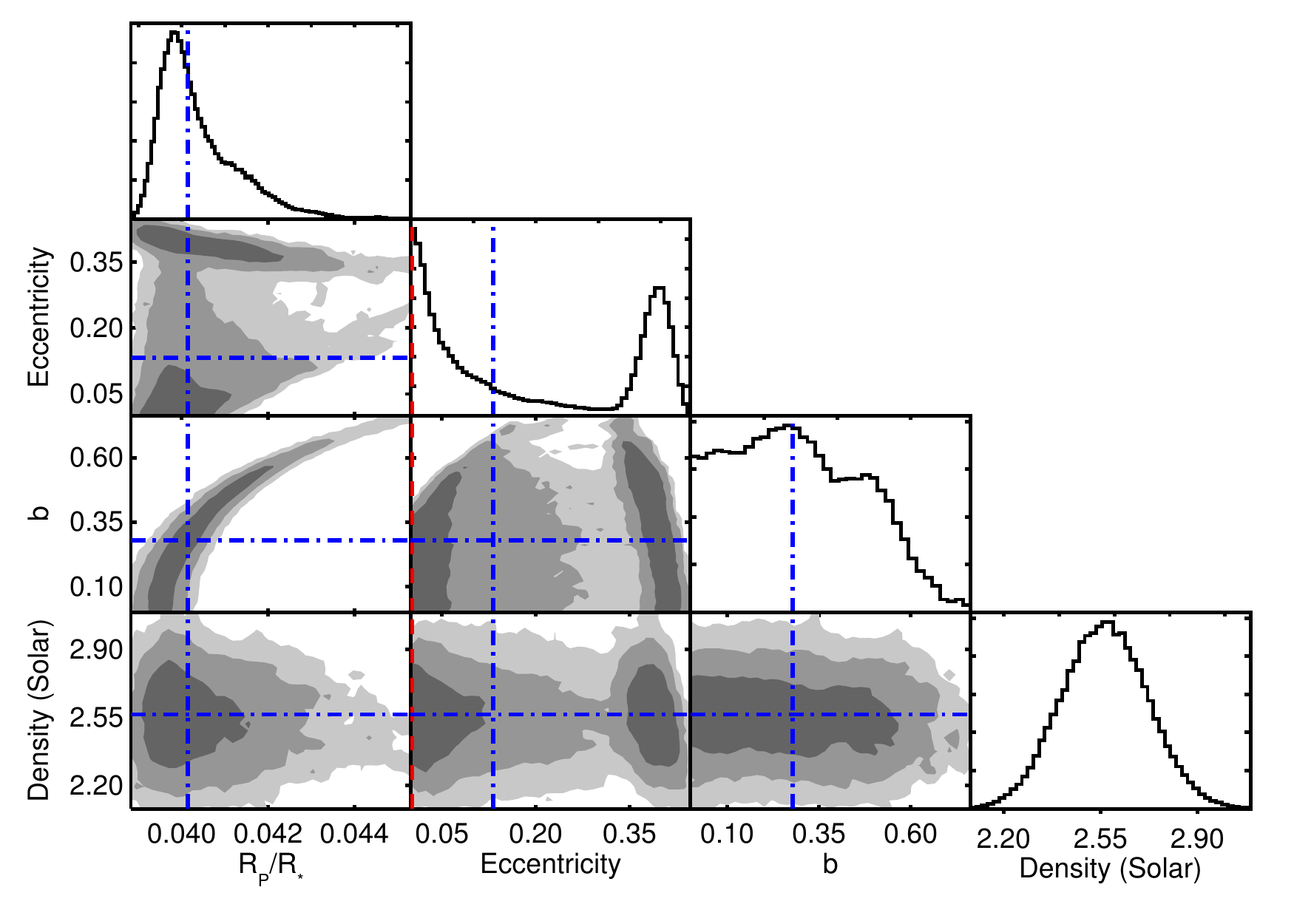}
\includegraphics[width=0.46\textwidth]{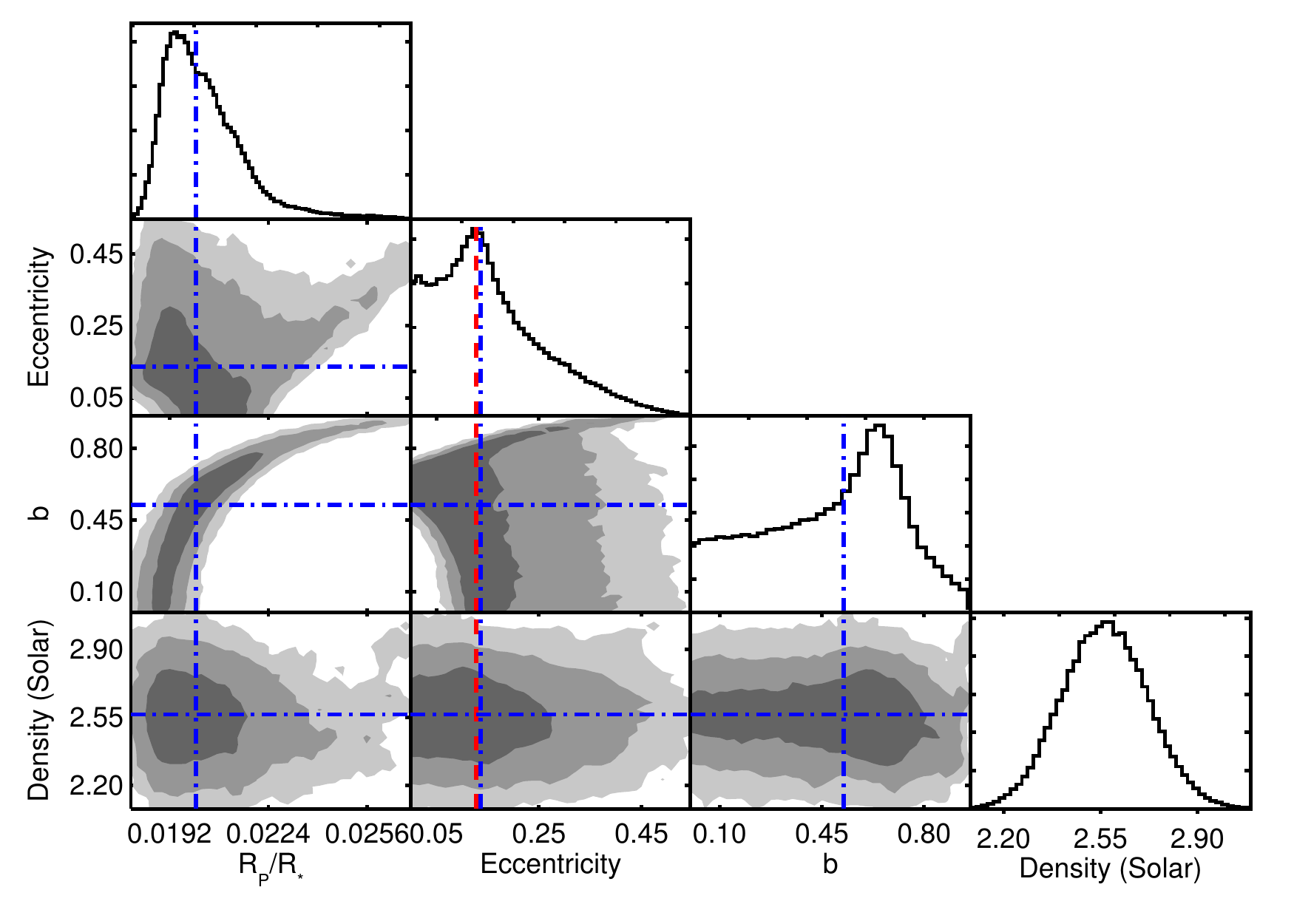}
\caption{Posteriors and correlations for a subset of the transit-fit parameters for planets b (top left), c (top right) and d (bottom). Dashed blue lines indicate the median of each distribution. Because the median is misleading for $e$ (due to the cutoff at $e=0$), we also show the mode as a red dashed line. Shaded regions (two-dimensional contours) refer to 68\%, 95\%, and 99.7\% of the points in the MCMC posterior. }
\label{fig:correlations}
\end{center}
\end{figure*}

We report the transit-fit parameters in Table~\ref{tab:planet}. For each parameter, we report the median value with the errors as the 84.1 and 15.9 percentile values (corresponding to 1$\sigma$ for one-dimensional Gaussian distributions). We show the distributions and correlations for a subset of parameters ($\rho_*$, $e$, $b$, and $R_P/R_*$) in Figure~\ref{fig:correlations}. 

The resulting eccentricities were generally small, as observed previously for small and multi-planet systems orbiting old stars \citep{Van-Eylen2015,Mann:2017ab}. The eccentricity posterior for the largest planet also appears bimodal, with peaks at $e\simeq0$ and $e\simeq0.4$. This is likely a consequence of the degeneracy between transit duration and $e$ for specific values of $\omega$ \citep[see][ for more information]{Mann:2017aa,Mann:2017ab}.

\begin{figure}[htbp]
\begin{center}
\includegraphics[width=0.48\textwidth]{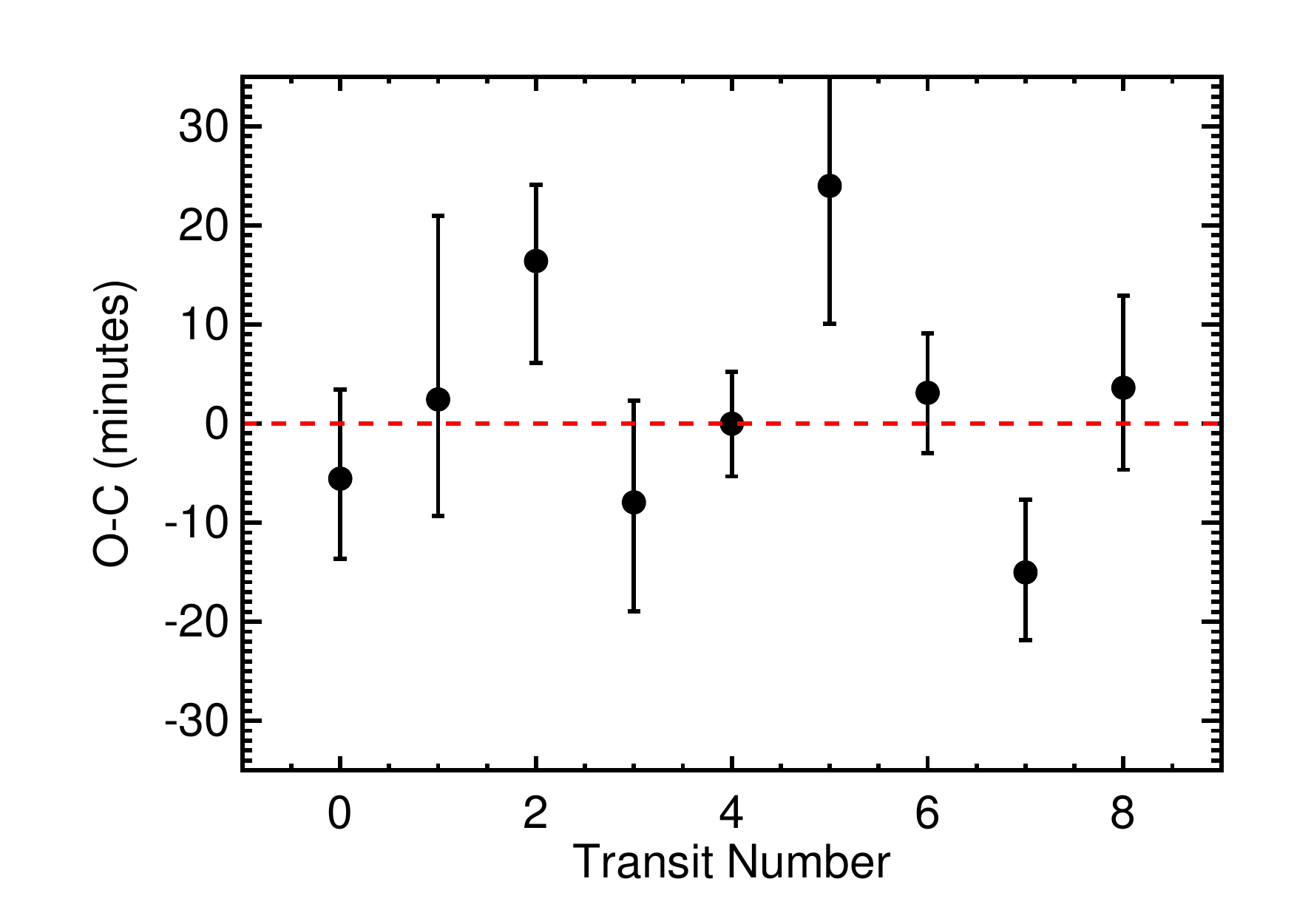}
\caption{Individual transit times for the shortest-period planet compared with the value expected from a linear ephemeris. Errors are from our MCMC analysis. Red dashed line marks no difference between the individual time and the expected linear ephemeris from our global fit.}
\label{fig:ttv}
\end{center}
\end{figure}

To keep the number of free parameters reasonable, our MCMC framework assumed a linear ephemeris for each planet. If transit timing variations (TTVs) are present, this spreads out the phase-folded transit, increasing the duration and decreasing the transit depth \citep[e.g.,][]{Swift2015}. To test for this we fit each transit event individually, following the procedure above, except applying a prior on all parameters derived from our simultaneous fit, with the exception $T_0$. We detected no significant TTV in any planet (e.g., Figure~\ref{fig:ttv}), although the constraints are rather weak. The longest-period planet has only three transits in the dataset, and the shortest-period planet is also the smallest, so individual transit times have errors of 5-20\,min. The expected TTV signal is sensitive to small changes in the exoplanet parameters \citep{Veras:2011nx}, but typical (detected) TTVs of similar-mass planets from {\it Kepler} are only 0.1-20\,min \citep[e.g.,][]{2013ApJS..208...16M, 2014ApJ...787...80H} except when near mean-motion resonance.

\section{False-Positive Analysis}\label{sec:false}
We assessed the probability that the transit signals are astrophysical false positives using the \textit{Vespa} software \citep{2015ascl.soft03011M}. \textit{Vespa} calculated the false-positive probability (FPP) by comparing the likelihood of a planet compared to that of three astrophysical false-positive scenarios; a background eclipsing binary, a bound eclipsing binary, and a hierarchical eclipsing system. The comparison was done using the transit shape and depth, properties of the star, and external constraints from ground-based imaging or our spectroscopy. As constraints, we included a contrast curve that we calculated from the 2MASS K-band image of \tar\ by fitting the star with a Moffat function and examining the fit residuals. 

For the \tar\ system, \textit{Vespa} returned very different false positive probabilities for the three individual planets. \textit{Vespa} calculated an FPP of 1.2\% for \tar\ b, 0.03\% for \tar\ c, and 10\% for \tar d. The different FPPs for the different planets reflect the large dynamic range in transit depth and signal-to-noise in the K2 light curves - the largest planet is easiest to validate because its transits are detected with high enough signal-to-noise to definitively show that the shape is flat-bottomed, while smaller planets more poorly constrained transit shapes. The \textit{Vespa} analysis does not take into account the fact that these candidates are found in a three-planet system. Candidates in multi-transiting systems are {\em a priori} much more likely to be real planets than single planets without any other transiting planets. \citet{2012ApJ...750..112L} estimate that candidates in systems with two other candidates are about 50 times more likely than average to be genuine exoplanets. When we apply this multiplicity boost, we find that all three planets in the \tar\ system are likely to be real - the FPPs drop to each less than one part in 500, sufficient for us to validate the planet candidates as genuine planets. 

\begin{figure}[htbp]
\begin{center}
\includegraphics[width=0.46\textwidth]{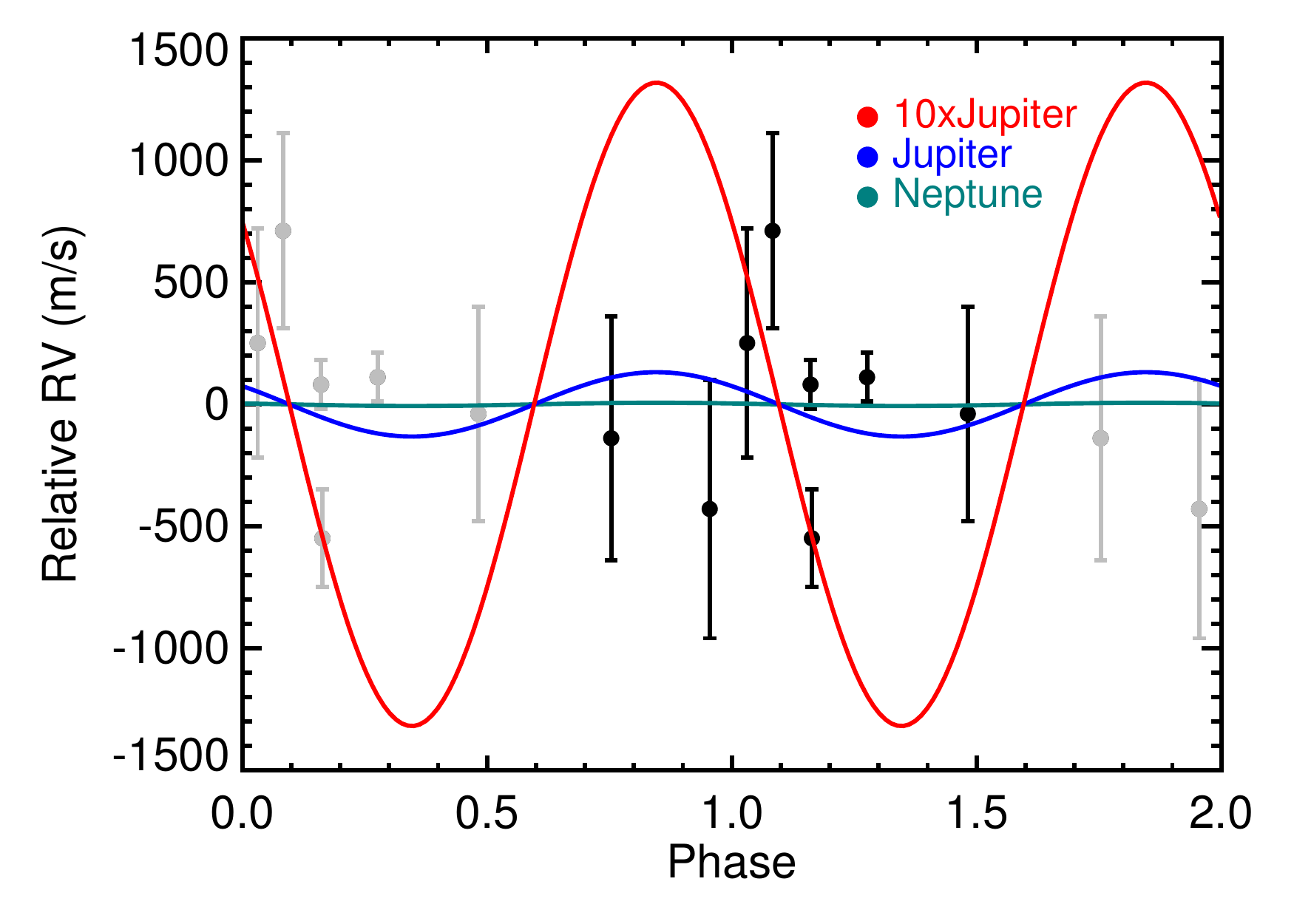}
\caption{Radial velocities of \tar, phase-folded to the largest planet. Grey points are repeated values so that two phases can be shown. The expected variation from a Neptune-mass (teal), Jupiter-mass (blue), and 10xJupiter (a low-mass brown dwarf; red) is overplotted for comparison, all assuming circular orbits. }
\label{fig:rvs}
\end{center}
\end{figure}

In addition to the constraints including in the \textit{Vespa} analysis, the three candidates all passed our standard pixel-level vetting tests, described by \citet{v16}, including searches for centroid motion in transit and for contamination from sources outside the photometric aperture (e.g., using different photometric apertures). The star also shows no detectable RV variability ($<0.5$\kms) over 30\,years of observations, ruling out most scenarios including a bound companion (Figure~\ref{fig:rvs}). No stars were detected within 20\arcsec\ of the source in SDSS or {\it Gaia} ($\Delta r>8$, $>2\arcsec$ from the source). This rules out widely separated background/foreground stars as the source of any transit signals, which has been a problem for earlier analyses \citep{2017arXiv170708007C}. None of these constraints are explicitly included in our \textit{Vespa} analysis, but significantly strengthen the likelihood that all three transits are planetary in origin.

\section{Discussion}\label{sec:discussion} 
Here we reported the discovery, confirmation, and characterization of a three-planet system in the Hyades cluster. Although this is now one of many planets discovered in young, nearby open clusters using {\it K2} data, it is the first multi-transiting system our survey has discovered. It also contains one of the youngest Earth-sized planets known \citep[depending the age assigned to Kepler-78, ][]{2013Natur.503..381H,2013Natur.503..377P}, offering the possibility of studying the history and evolution of Earth-sized planets.

Multi-transiting systems offer a wider set of science applications than their single-planet counterparts. Eccentricities of multi-transiting systems can be measured from the light curve alone, using the comparison of expected versus observed transit durations between planets in the same system \citep[e.g.,][]{2012MNRAS.421.1166K}. In the case where the stellar density is independently known, multiple transits with varying impact parameters can provide extremely strong constraints on the eccentricities of the planets \citep[e.g.,][]{Van-Eylen2015}. In the case the eccentricity is known these can be powerful tests of stellar parameters \citep[e.g.,][]{Mann:2017ab}. For \tar, it may be possible to constrain the eccentricities using dynamical stability arguments, especially when coupled with the lack of TTVs. However, \ktwo\ long-cadence data are prohibitive for these tasks, since the primary constraint comes from the transit duration \citep{Seager:2003lr}, and the ingress/egress is often shorter than the integration time. Fortunately, the largest planet is well within the reach of ground-based follow-up, and the smaller planets using space-based observatories, offering the chance for significantly higher-cadence observations. 

Multi-transiting systems also facilitate the measurement of masses through TTVs \citep[e.g.,][]{Hadden:2017aa}. There is no evidence for TTVs in any of the planets (Section~\ref{sec:transit}), however, the baseline is too small for any but the shortest-period planet (e.g., there are only three transits of the longest-period planet). This is also the smallest planet and has the least-well measured individual transit times (errors of 5-20\,min). In addition to exploring the eccentricities and stellar parameters, follow-up at higher cadence could be useful to detect any TTV signals.

\subsection{Prospects for Planet Masses}
Thanks to its bright V-band magnitude and relatively low stellar variability, \tar\ is a highly promising target for precise radial velocity follow-up to for the first time measure the masses of small transiting planets in an open cluster. Of the {\it K2} young planets found thus far, only one other host star besides \tar\ is bright enough ($V\lesssim12$) for high-precision RVs with existing instrumentation: K2-100b, a Neptune-sized planet around an F dwarf in Praesepe, with an expected RV signal of 4\ms. However, K2-100 has a high projected rotational velocity of about 15 \kms, making precise RV measurements more difficult, and shows high-amplitude ($\approx 1.5\%$ peak to peak) stellar variability that changes in pattern and amplitude on $\simeq$week timescales, likely generating complex stellar signals of with amplitudes of about 100 \ms. A precise mass determination for K2-100 would therefore be quite difficult and observationally expensive. 

\tar\ also shows variability in its light curve and longer-term fluctuations in this variability, as expected for a young late-K dwarf. However, \tar\ is on the low end of the activity scale for Hyades members, both in terms of its rotation period (Figure~\ref{fig:rot}) and variability amplitude (Figure~\ref{fig:lc}). While the actual relation between photometric and RV variability amplitude is complex \citep[e.g.,][]{2012MNRAS.419.3147A,Oshagh:2017aa}, in the case of a spot dominated atmosphere we can approximate the RV variability as $\sim\sigma_\mathrm{phot}$*\vsini, where $\sigma_\mathrm{phot}$ is the photometric variability in the same band. While most of the ZEIT planet hosts have $\sigma_\mathrm{phot}$ of 2-5\%, the value for \tar\ is $\simeq$0.3\%, suggesting RV variability of only $\sim$6-9\,m/s. 

We can compare this value to the expected RV signal from the planets themselves. Using the $M_*-R_*$ and $\rho_*-R_*$ relations from \citet{2014ApJ...783L...6W}, we estimate masses of $\sim$ 1$M_\oplus$, 7$M_\oplus$, and 4$M_\oplus$ for planets b, c, and d. This corresponds to RV amplitudes of $\sim$ 2\ms, 1\ms, and 0.4\ms. While all three are below the expected RV signal from the star, we benefit from prior knowledge about the star's rotation period and the orbital periods of the planets. This, combined with simultaneous optical monitoring puts mass measurements of at least two of these planets within reach of the highest precision RV spectrographs \citep[e.g.,][]{2015ApJ...814L..21D,2015A&A...584A..72M}.

\acknowledgements
During the writing of this paper we became aware of a parallel analysis of the same system, \citet{2017arXiv170910398C}. We thank David Ciardi and collaborators for agreeing to coordinate paper submission. We also note another paper appeared after the submission of our paper, \citet{2017arXiv171007203L} that is broadly consistent with our own results. 

We thank the anonymous referee for their fast and thoughtful report. 

AWM was supported through Hubble Fellowship grant 51364 awarded by the Space Telescope Science Institute, which is operated by the Association of Universities for Research in Astronomy, Inc., for NASA, under contract NAS 5-26555. ACR was supported (in part) by NASA K2 Guest Observer Cycle 4 grant NNX17AF71G. This work was performed in part under contract with the California Institute of Technology/Jet Propulsion Laboratory funded by NASA through the Sagan Fellowship Program executed by the NASA Exoplanet Science Institute.

This work used the Immersion Grating Infrared Spectrograph (IGRINS) that was developed under a collaboration between the University of Texas at Austin and the Korea Astronomy and Space Science Institute (KASI) with the financial support of the US National Science Foundation under grant AST-1229522, of the University of Texas at Austin, and of the Korean GMT Project of KASI. The IGRINS pipeline package PLP was developed by Dr. Jae-Joon Lee at Korea Astronomy and Space Science Institute and Professor Soojong Pak's team at Kyung Hee University.  STScI is operated by the Association of Universities for Research in Astronomy, Inc., under NASA contract NAS5-26555. Support for MAST for non-HST data is provided by the NASA Office of Space Science via grant NNX09AF08G and by other grants and contracts. This research was made possible through the use of the AAVSO Photometric All-Sky Survey (APASS), funded by the Robert Martin Ayers Sciences Fund. The authors acknowledge the Texas Advanced Computing Center (TACC) at The University of Texas at Austin for providing HPC resources that have contributed to the research results reported within this paper\footnote{\href{http://www.tacc.utexas.edu}{http://www.tacc.utexas.edu}}. These results made use of the Discovery Channel Telescope at Lowell Observatory. Lowell is a private, non-profit institution dedicated to astrophysical research and public appreciation of astronomy and operates the DCT in partnership with Boston University, the University of Maryland, the University of Toledo, Northern Arizona University and Yale University.

\facilities{FLWO:1.5m (TRES, Digital Speedometers), DCT (IGRINS), Kepler, CTIO:2MASS, Sloan, Gaia}

\software{vespa \citep{2015ascl.soft03011M}, emcee \citep{Foreman-Mackey2013}, astropy \citep{2013A&A...558A..33A}}

\bibliography{fullbiblio.bib}

\begin{thebibliography}{}
\expandafter\ifx\csname natexlab\endcsname\relax\def\natexlab#1{#1}\fi

\bibitem[{{Ahn} {et~al.}(2012){Ahn}, {Alexandroff}, {Allende Prieto},
  {Anderson}, {Anderton}, {Andrews}, {Aubourg}, {Bailey}, {Balbinot}, {Barnes},
  \& et~al.}]{Ahn:2012kx}
{Ahn}, C.~P., {Alexandroff}, R., {Allende Prieto}, C., {et~al.} 2012, \apjs,
  203, 21

\bibitem[{{Aigrain} {et~al.}(2012){Aigrain}, {Pont}, \&
  {Zucker}}]{2012MNRAS.419.3147A}
{Aigrain}, S., {Pont}, F., \& {Zucker}, S. 2012, \mnras, 419, 3147

\bibitem[{{Allard} {et~al.}(2011){Allard}, {Homeier}, \&
  {Freytag}}]{Allard2011}
{Allard}, F., {Homeier}, D., \& {Freytag}, B. 2011, in Astronomical Society of
  the Pacific Conference Series, Vol. 448, 16th Cambridge Workshop on Cool
  Stars, Stellar Systems, and the Sun, ed. C.~{Johns-Krull}, M.~K. {Browning},
  \& A.~A. {West}, 91

\bibitem[{{Altmann} {et~al.}(2017){Altmann}, {Roeser}, {Demleitner}, {Bastian},
  \& {Schilbach}}]{hsoy}
{Altmann}, M., {Roeser}, S., {Demleitner}, M., {Bastian}, U., \& {Schilbach},
  E. 2017, \aap, 600, L4

\bibitem[{{Ansdell} {et~al.}(2016){Ansdell}, {Gaidos}, {Rappaport}, {Jacobs},
  {LaCourse}, {Jek}, {Mann}, {Wyatt}, {Kennedy}, {Williams}, \&
  {Boyajian}}]{2016ApJ...816...69A}
{Ansdell}, M., {Gaidos}, E., {Rappaport}, S.~A., {et~al.} 2016, \apj, 816, 69

\bibitem[{{Astropy Collaboration} {et~al.}(2013){Astropy Collaboration},
  {Robitaille}, {Tollerud}, {Greenfield}, {Droettboom}, {Bray}, {Aldcroft},
  {Davis}, {Ginsburg}, {Price-Whelan}, {Kerzendorf}, {Conley}, {Crighton},
  {Barbary}, {Muna}, {Ferguson}, {Grollier}, {Parikh}, {Nair}, {Unther},
  {Deil}, {Woillez}, {Conseil}, {Kramer}, {Turner}, {Singer}, {Fox}, {Weaver},
  {Zabalza}, {Edwards}, {Azalee Bostroem}, {Burke}, {Casey}, {Crawford},
  {Dencheva}, {Ely}, {Jenness}, {Labrie}, {Lim}, {Pierfederici}, {Pontzen},
  {Ptak}, {Refsdal}, {Servillat}, \& {Streicher}}]{2013A&A...558A..33A}
{Astropy Collaboration}, {Robitaille}, T.~P., {Tollerud}, E.~J., {et~al.} 2013,
  \aap, 558, A33

\bibitem[{{Baraffe} {et~al.}(2015){Baraffe}, {Homeier}, {Allard}, \&
  {Chabrier}}]{BHAC15}
{Baraffe}, I., {Homeier}, D., {Allard}, F., \& {Chabrier}, G. 2015, \aap, 577,
  A42

\bibitem[{{Bastien} {et~al.}(2014){Bastien}, {Stassun}, \&
  {Pepper}}]{Bastien2014}
{Bastien}, F.~A., {Stassun}, K.~G., \& {Pepper}, J. 2014, \apjl, 788, L9

\bibitem[{{Becker} {et~al.}(2015){Becker}, {Vanderburg}, {Adams}, {Rappaport},
  \& {Schwengeler}}]{Becker2015}
{Becker}, J.~C., {Vanderburg}, A., {Adams}, F.~C., {Rappaport}, S.~A., \&
  {Schwengeler}, H.~M. 2015, \apjl, 812, L18

\bibitem[{{Borucki} {et~al.}(2010){Borucki}, {Koch}, {Basri}, {Batalha},
  {Brown}, {Caldwell}, {Caldwell}, {Christensen-Dalsgaard}, {Cochran},
  {DeVore}, {Dunham}, {Dupree}, {Gautier}, {Geary}, {Gilliland}, {Gould},
  {Howell}, {Jenkins}, {Kondo}, {Latham}, {Marcy}, {Meibom}, {Kjeldsen},
  {Lissauer}, {Monet}, {Morrison}, {Sasselov}, {Tarter}, {Boss}, {Brownlee},
  {Owen}, {Buzasi}, {Charbonneau}, {Doyle}, {Fortney}, {Ford}, {Holman},
  {Seager}, {Steffen}, {Welsh}, {Rowe}, {Anderson}, {Buchhave}, {Ciardi},
  {Walkowicz}, {Sherry}, {Horch}, {Isaacson}, {Everett}, {Fischer}, {Torres},
  {Johnson}, {Endl}, {MacQueen}, {Bryson}, {Dotson}, {Haas}, {Kolodziejczak},
  {Van Cleve}, {Chandrasekaran}, {Twicken}, {Quintana}, {Clarke}, {Allen},
  {Li}, {Wu}, {Tenenbaum}, {Verner}, {Bruhweiler}, {Barnes}, \&
  {Prsa}}]{Borucki2010}
{Borucki}, W.~J., {Koch}, D., {Basri}, G., {et~al.} 2010, Science, 327, 977

\bibitem[{{Brandt} \& {Huang}(2015)}]{Brandt2015}
{Brandt}, T.~D., \& {Huang}, C.~X. 2015, \apj, 807, 24

\bibitem[{{Buchhave} {et~al.}(2012){Buchhave}, {Latham}, {Johansen},
  {Bizzarro}, {Torres}, {Rowe}, {Batalha}, {Borucki}, {Brugamyer}, {Caldwell},
  {Bryson}, {Ciardi}, {Cochran}, {Endl}, {Esquerdo}, {Ford}, {Geary},
  {Gilliland}, {Hansen}, {Isaacson}, {Laird}, {Lucas}, {Marcy}, {Morse},
  {Robertson}, {Shporer}, {Stefanik}, {Still}, \&
  {Quinn}}]{2012Natur.486..375B}
{Buchhave}, L.~A., {Latham}, D.~W., {Johansen}, A., {et~al.} 2012, \nat, 486,
  375

\bibitem[{{Buchhave} {et~al.}(2014){Buchhave}, {Bizzarro}, {Latham},
  {Sasselov}, {Cochran}, {Endl}, {Isaacson}, {Juncher}, \&
  {Marcy}}]{2014Natur.509..593B}
{Buchhave}, L.~A., {Bizzarro}, M., {Latham}, D.~W., {et~al.} 2014, \nat, 509,
  593

\bibitem[{{Cabrera} {et~al.}(2017){Cabrera}, {Barros}, {Armstrong}, {Hidalgo},
  {Santos}, {Almenara}, {Alonso}, {Deleuil}, {Demangeon}, {Diaz}, {Lendl},
  {Pfaff}, {Rauer}, {Santerne}, {Serrano}, \& {Zucker}}]{2017arXiv170708007C}
{Cabrera}, J., {Barros}, S.~C.~C., {Armstrong}, D., {et~al.} 2017, ArXiv
  e-prints, arXiv:1707.08007

\bibitem[{{Choi} {et~al.}(2016){Choi}, {Dotter}, {Conroy}, {Cantiello},
  {Paxton}, \& {Johnson}}]{2016ApJ...823..102C}
{Choi}, J., {Dotter}, A., {Conroy}, C., {et~al.} 2016, \apj, 823, 102

\bibitem[{{Christiansen} {et~al.}(2016){Christiansen}, {Clarke}, {Burke},
  {Jenkins}, {Bryson}, {Coughlin}, {Mullally}, {Thompson}, {Twicken},
  {Batalha}, {Haas}, {Catanzarite}, {Campbell}, {Kamal Uddin}, {Zamudio},
  {Smith}, \& {Henze}}]{2016ApJ...828...99C}
{Christiansen}, J.~L., {Clarke}, B.~D., {Burke}, C.~J., {et~al.} 2016, \apj,
  828, 99

\bibitem[{{Ciardi} {et~al.}(2017){Ciardi}, {Crossfield}, {Feinstein},
  {Schlieder}, {Petigura}, {David}, {Bristow}, {Patel}, {Arnold}, {Benneke},
  {Christiansen}, {Dressing}, {Fulton}, {Howard}, {Isaacson}, {Sinukoff}, \&
  {Thackeray}}]{2017arXiv170910398C}
{Ciardi}, D.~R., {Crossfield}, I.~J.~M., {Feinstein}, A.~D., {et~al.} 2017,
  ArXiv e-prints, arXiv:1709.10398

\bibitem[{{Cohen} {et~al.}(2003){Cohen}, {Wheaton}, \&
  {Megeath}}]{2003AJ....126.1090C}
{Cohen}, M., {Wheaton}, W.~A., \& {Megeath}, S.~T. 2003, \aj, 126, 1090

\bibitem[{{Crossfield} {et~al.}(2016){Crossfield}, {Ciardi}, {Petigura},
  {Sinukoff}, {Schlieder}, {Howard}, {Beichman}, {Isaacson}, {Dressing},
  {Christiansen}, {Fulton}, {L{\'e}pine}, {Weiss}, {Hirsch}, {Livingston},
  {Baranec}, {Law}, {Riddle}, {Ziegler}, {Howell}, {Horch}, {Everett}, {Teske},
  {Martinez}, {Obermeier}, {Benneke}, {Scott}, {Deacon}, {Aller}, {Hansen},
  {Mancini}, {Ciceri}, {Brahm}, {Jord{\'a}n}, {Knutson}, {Henning}, {Bonnefoy},
  {Liu}, {Crepp}, {Lothringer}, {Hinz}, {Bailey}, {Skemer}, \&
  {Defrere}}]{Crossfield:2016aa}
{Crossfield}, I.~J.~M., {Ciardi}, D.~R., {Petigura}, E.~A., {et~al.} 2016,
  \apjs, 226, 7

\bibitem[{{Dahm}(2015)}]{2015ApJ...813..108D}
{Dahm}, S.~E. 2015, \apj, 813, 108

\bibitem[{{David} {et~al.}(2016){David}, {Hillenbrand}, {Petigura},
  {Carpenter}, {Crossfield}, {Hinkley}, {Ciardi}, {Howard}, {Isaacson}, {Cody},
  {Schlieder}, {Beichman}, \& {Barenfeld}}]{David2016b}
{David}, T.~J., {Hillenbrand}, L.~A., {Petigura}, E.~A., {et~al.} 2016, \nat,
  534, 658

\bibitem[{{de Bruijne}(2012)}]{2012Ap&SS.341...31D}
{de Bruijne}, J.~H.~J. 2012, \apss, 341, 31

\bibitem[{{Donati} {et~al.}(2017){Donati}, {Yu}, {Moutou}, {Cameron}, {Malo},
  {Grankin}, {H{\'e}brard}, {Hussain}, {Vidotto}, {Alencar}, {Haywood},
  {Bouvier}, {Petit}, {Takami}, {Herczeg}, {Gregory}, {Jardine}, \&
  {Morin}}]{Donati:2017aa}
{Donati}, J.-F., {Yu}, L., {Moutou}, C., {et~al.} 2017, \mnras, 465, 3343

\bibitem[{{Dotter}(2016)}]{2016ApJS..222....8D}
{Dotter}, A. 2016, \apjs, 222, 8

\bibitem[{{Douglas} {et~al.}(2016){Douglas}, {Ag{\"u}eros}, {Covey}, {Cargile},
  {Barclay}, {Cody}, {Howell}, \& {Kopytova}}]{2016ApJ...822...47D}
{Douglas}, S.~T., {Ag{\"u}eros}, M.~A., {Covey}, K.~R., {et~al.} 2016, \apj,
  822, 47

\bibitem[{{Douglas} {et~al.}(2017){Douglas}, {Ag{\"u}eros}, {Covey}, \&
  {Kraus}}]{2017ApJ...842...83D}
{Douglas}, S.~T., {Ag{\"u}eros}, M.~A., {Covey}, K.~R., \& {Kraus}, A. 2017,
  \apj, 842, 83

\bibitem[{{Dressing} \& {Charbonneau}(2013)}]{Dressing2013}
{Dressing}, C.~D., \& {Charbonneau}, D. 2013, \apj, 767, 95

\bibitem[{{Dressing} {et~al.}(2017){Dressing}, {Newton}, {Schlieder},
  {Charbonneau}, {Knutson}, {Vanderburg}, \& {Sinukoff}}]{Dressing2017}
{Dressing}, C.~D., {Newton}, E.~R., {Schlieder}, J.~E., {et~al.} 2017, \apj,
  836, 167

\bibitem[{{Dumusque} {et~al.}(2015){Dumusque}, {Glenday}, {Phillips},
  {Buchschacher}, {Collier Cameron}, {Cecconi}, {Charbonneau}, {Cosentino},
  {Ghedina}, {Latham}, {Li}, {Lodi}, {Lovis}, {Molinari}, {Pepe}, {Udry},
  {Sasselov}, {Szentgyorgyi}, \& {Walsworth}}]{2015ApJ...814L..21D}
{Dumusque}, X., {Glenday}, A., {Phillips}, D.~F., {et~al.} 2015, \apjl, 814,
  L21

\bibitem[{{Dutra-Ferreira} {et~al.}(2016){Dutra-Ferreira}, {Pasquini},
  {Smiljanic}, {Porto de Mello}, \& {Steffen}}]{Dutra-Ferreira2016}
{Dutra-Ferreira}, L., {Pasquini}, L., {Smiljanic}, R., {Porto de Mello}, G.~F.,
  \& {Steffen}, M. 2016, \aap, 585, A75

\bibitem[{{Ehrenreich} \& {D{\'e}sert}(2011)}]{2011A&A...529A.136E}
{Ehrenreich}, D., \& {D{\'e}sert}, J.-M. 2011, \aap, 529, A136

\bibitem[{{Ehrenreich} {et~al.}(2015){Ehrenreich}, {Bourrier}, {Wheatley}, {Des
  Etangs}, {H{\'e}brard}, {Udry}, {Bonfils}, {Delfosse}, {D{\'e}sert}, {Sing},
  \& {Vidal-Madjar}}]{Ehrenreich2015}
{Ehrenreich}, D., {Bourrier}, V., {Wheatley}, P.~J., {et~al.} 2015, \nat, 522,
  459

\bibitem[{{Foreman-Mackey} {et~al.}(2013){Foreman-Mackey}, {Hogg}, {Lang}, \&
  {Goodman}}]{Foreman-Mackey2013}
{Foreman-Mackey}, D., {Hogg}, D.~W., {Lang}, D., \& {Goodman}, J. 2013, \pasp,
  125, 306

\bibitem[{{Fressin} {et~al.}(2013){Fressin}, {Torres}, {Charbonneau}, {Bryson},
  {Christiansen}, {Dressing}, {Jenkins}, {Walkowicz}, \&
  {Batalha}}]{Fressin:2013qy}
{Fressin}, F., {Torres}, G., {Charbonneau}, D., {et~al.} 2013, \apj, 766, 81

\bibitem[{{Gaia Collaboration} {et~al.}(2016){Gaia Collaboration}, {Brown},
  {Vallenari}, {Prusti}, {de Bruijne}, {Mignard}, {Drimmel}, {Babusiaux},
  {Bailer-Jones}, {Bastian}, \& et~al.}]{gaiadr1}
{Gaia Collaboration}, {Brown}, A.~G.~A., {Vallenari}, A., {et~al.} 2016, \aap,
  595, A2

\bibitem[{{Gaidos} \& {Mann}(2013)}]{Gaidos2013}
{Gaidos}, E., \& {Mann}, A.~W. 2013, \apj, 762, 41

\bibitem[{{Gaidos} {et~al.}(2016){Gaidos}, {Mann}, {Kraus}, \&
  {Ireland}}]{Gaidos2016b}
{Gaidos}, E., {Mann}, A.~W., {Kraus}, A.~L., \& {Ireland}, M. 2016, \mnras,
  457, 2877

\bibitem[{{Gaidos} {et~al.}(2014){Gaidos}, {Anderson}, {L{\'e}pine},
  {Col{\'o}n}, {Maravelias}, {Narita}, {Chang}, {Beyer}, {Fukui}, {Armstrong},
  {Zezas}, {Fulton}, {Mann}, {West}, \& {Faedi}}]{2014MNRAS.437.3133G}
{Gaidos}, E., {Anderson}, D.~R., {L{\'e}pine}, S., {et~al.} 2014, \mnras, 437,
  3133

\bibitem[{{Gaidos} {et~al.}(2017){Gaidos}, {Mann}, {Rizzuto}, {Nofi}, {Mace},
  {Vanderburg}, {Feiden}, {Narita}, {Takeda}, {Esposito}, {De Rosa}, {Ansdell},
  {Hirano}, {Graham}, {Kraus}, \& {Jaffe}}]{Gaidos:2017aa}
{Gaidos}, E., {Mann}, A.~W., {Rizzuto}, A., {et~al.} 2017, \mnras, 464, 850

\bibitem[{{Gillen} {et~al.}(2017){Gillen}, {Hillenbrand}, {David}, {Aigrain},
  {Rebull}, {Stauffer}, {Cody}, \& {Queloz}}]{2017arXiv170603084G}
{Gillen}, E., {Hillenbrand}, L.~A., {David}, T.~J., {et~al.} 2017, ArXiv
  e-prints, arXiv:1706.03084

\bibitem[{{Goldman} {et~al.}(2013){Goldman}, {R{\"o}ser}, {Schilbach},
  {Magnier}, {Olczak}, {Henning}, {Juri{\'c}}, {Schlafly}, {Chen}, {Platais},
  {Burgett}, {Hodapp}, {Heasley}, {Kudritzki}, {Morgan}, {Price}, {Tonry}, \&
  {Wainscoat}}]{2013A&A...559A..43G}
{Goldman}, B., {R{\"o}ser}, S., {Schilbach}, E., {et~al.} 2013, \aap, 559, A43

\bibitem[{{Guo} {et~al.}(2017){Guo}, {Johnson}, {Mann}, {Kraus}, {Curtis}, \&
  {Latham}}]{2017ApJ...838...25G}
{Guo}, X., {Johnson}, J.~A., {Mann}, A.~W., {et~al.} 2017, \apj, 838, 25

\bibitem[{{Hadden} \& {Lithwick}(2014)}]{2014ApJ...787...80H}
{Hadden}, S., \& {Lithwick}, Y. 2014, \apj, 787, 80

\bibitem[{{Hadden} \& {Lithwick}(2017)}]{Hadden:2017aa}
---. 2017, \aj, 154, 5

\bibitem[{{Henden} {et~al.}(2012){Henden}, {Levine}, {Terrell}, {Smith}, \&
  {Welch}}]{Henden:2012fk}
{Henden}, A.~A., {Levine}, S.~E., {Terrell}, D., {Smith}, T.~C., \& {Welch}, D.
  2012, Journal of the American Association of Variable Star Observers
  (JAAVSO), 40, 430

\bibitem[{{Hermes} {et~al.}(2017){Hermes}, {G{\"a}nsicke}, {Gentile Fusillo},
  {Raddi}, {Hollands}, {Dennihy}, {Fuchs}, \& {Redfield}}]{2017MNRAS.468.1946H}
{Hermes}, J.~J., {G{\"a}nsicke}, B.~T., {Gentile Fusillo}, N.~P., {et~al.}
  2017, \mnras, 468, 1946

\bibitem[{{Howard} {et~al.}(2013){Howard}, {Sanchis-Ojeda}, {Marcy}, {Johnson},
  {Winn}, {Isaacson}, {Fischer}, {Fulton}, {Sinukoff}, \&
  {Fortney}}]{2013Natur.503..381H}
{Howard}, A.~W., {Sanchis-Ojeda}, R., {Marcy}, G.~W., {et~al.} 2013, \nat, 503,
  381

\bibitem[{{Howell} {et~al.}(2014){Howell}, {Sobeck}, {Haas}, {Still},
  {Barclay}, {Mullally}, {Troeltzsch}, {Aigrain}, {Bryson}, {Caldwell},
  {Chaplin}, {Cochran}, {Huber}, {Marcy}, {Miglio}, {Najita}, {Smith},
  {Twicken}, \& {Fortney}}]{Howell2014}
{Howell}, S.~B., {Sobeck}, C., {Haas}, M., {et~al.} 2014, \pasp, 126, 398

\bibitem[{{Huber} {et~al.}(2014){Huber}, {Silva Aguirre}, {Matthews},
  {Pinsonneault}, {Gaidos}, {Garc{\'{\i}}a}, {Hekker}, {Mathur}, {Mosser},
  {Torres}, {Bastien}, {Basu}, {Bedding}, {Chaplin}, {Demory}, {Fleming},
  {Guo}, {Mann}, {Rowe}, {Serenelli}, {Smith}, \&
  {Stello}}]{2014ApJS..211....2H}
{Huber}, D., {Silva Aguirre}, V., {Matthews}, J.~M., {et~al.} 2014, \apjs, 211,
  2

\bibitem[{{Huber} {et~al.}(2016){Huber}, {Bryson}, {Haas}, {Barclay},
  {Barentsen}, {Howell}, {Sharma}, {Stello}, \& {Thompson}}]{Huber2016}
{Huber}, D., {Bryson}, S.~T., {Haas}, M.~R., {et~al.} 2016, \apjs, 224, 2

\bibitem[{{Husser} {et~al.}(2013){Husser}, {Wende-von Berg}, {Dreizler},
  {Homeier}, {Reiners}, {Barman}, \& {Hauschildt}}]{2013A&A...553A...6H}
{Husser}, T.-O., {Wende-von Berg}, S., {Dreizler}, S., {et~al.} 2013, \aap,
  553, A6

\bibitem[{{Jenkins} {et~al.}(2010{\natexlab{a}}){Jenkins}, {Caldwell},
  {Chandrasekaran}, {Twicken}, {Bryson}, {Quintana}, {Clarke}, {Li}, {Allen},
  {Tenenbaum}, {Wu}, {Klaus}, {Van Cleve}, {Dotson}, {Haas}, {Gilliland},
  {Koch}, \& {Borucki}}]{Jenkins:2010qy}
{Jenkins}, J.~M., {Caldwell}, D.~A., {Chandrasekaran}, H., {et~al.}
  2010{\natexlab{a}}, \apjl, 713, L120

\bibitem[{{Jenkins} {et~al.}(2010{\natexlab{b}}){Jenkins}, {Caldwell},
  {Chandrasekaran}, {Twicken}, {Bryson}, {Quintana}, {Clarke}, {Li}, {Allen},
  {Tenenbaum}, {Wu}, {Klaus}, {Middour}, {Cote}, {McCauliff}, {Girouard},
  {Gunter}, {Wohler}, {Sommers}, {Hall}, {Uddin}, {Wu}, {Bhavsar}, {Van Cleve},
  {Pletcher}, {Dotson}, {Haas}, {Gilliland}, {Koch}, \&
  {Borucki}}]{2010ApJ...713L..87J}
---. 2010{\natexlab{b}}, \apjl, 713, L87

\bibitem[{{Kipping}(2010)}]{Kipping:2010lr}
{Kipping}, D.~M. 2010, \mnras, 408, 1758

\bibitem[{{Kipping}(2013)}]{Kipping2013}
---. 2013, \mnras, 435, 2152

\bibitem[{{Kipping} {et~al.}(2012){Kipping}, {Dunn}, {Jasinski}, \&
  {Manthri}}]{2012MNRAS.421.1166K}
{Kipping}, D.~M., {Dunn}, W.~R., {Jasinski}, J.~M., \& {Manthri}, V.~P. 2012,
  \mnras, 421, 1166

\bibitem[{{Kov{\'a}cs} {et~al.}(2002){Kov{\'a}cs}, {Zucker}, \&
  {Mazeh}}]{Kovacs2002}
{Kov{\'a}cs}, G., {Zucker}, S., \& {Mazeh}, T. 2002, \aap, 391, 369

\bibitem[{{Kraus} {et~al.}(2017{\natexlab{a}}){Kraus}, {Herczeg}, {Rizzuto},
  {Mann}, {Slesnick}, {Carpenter}, {Hillenbrand}, \&
  {Mamajek}}]{2017ApJ...838..150K}
{Kraus}, A.~L., {Herczeg}, G.~J., {Rizzuto}, A.~C., {et~al.}
  2017{\natexlab{a}}, \apj, 838, 150

\bibitem[{{Kraus} {et~al.}(2017{\natexlab{b}}){Kraus}, {Douglas}, {Mann},
  {Ag{\"u}eros}, {Law}, {Covey}, {Feiden}, {Rizzuto}, {Howard}, {Isaacson},
  {Gaidos}, {Torres}, \& {Bakos}}]{2017ApJ...845...72K}
{Kraus}, A.~L., {Douglas}, S.~T., {Mann}, A.~W., {et~al.} 2017{\natexlab{b}},
  \apj, 845, 72

\bibitem[{{Kreidberg}(2015)}]{Kreidberg2015}
{Kreidberg}, L. 2015, \pasp, 127, 1161

\bibitem[{{Kurucz}(1992)}]{kurucz1992}
{Kurucz}, R.~L. 1992, in IAU Symposium, Vol. 149, The Stellar Populations of
  Galaxies, ed. B.~{Barbuy} \& A.~{Renzini}, 225

\bibitem[{Lee(2015)}]{IGRINS_plp}
Lee, J.-J. 2015, plp: Version 2.0, doi:10.5281/zenodo.18579

\bibitem[{{L{\'e}pine} {et~al.}(2013){L{\'e}pine}, {Hilton}, {Mann}, {Wilde},
  {Rojas-Ayala}, {Cruz}, \& {Gaidos}}]{Lepine:2013}
{L{\'e}pine}, S., {Hilton}, E.~J., {Mann}, A.~W., {et~al.} 2013, \aj, 145, 102

\bibitem[{{Linsky} {et~al.}(2010){Linsky}, {Yang}, {France}, {Froning},
  {Green}, {Stocke}, \& {Osterman}}]{2010ApJ...717.1291L}
{Linsky}, J.~L., {Yang}, H., {France}, K., {et~al.} 2010, \apj, 717, 1291

\bibitem[{{Lissauer} {et~al.}(2012){Lissauer}, {Marcy}, {Rowe}, {Bryson},
  {Adams}, {Buchhave}, {Ciardi}, {Cochran}, {Fabrycky}, {Ford}, {Fressin},
  {Geary}, {Gilliland}, {Holman}, {Howell}, {Jenkins}, {Kinemuchi}, {Koch},
  {Morehead}, {Ragozzine}, {Seader}, {Tanenbaum}, {Torres}, \&
  {Twicken}}]{2012ApJ...750..112L}
{Lissauer}, J.~J., {Marcy}, G.~W., {Rowe}, J.~F., {et~al.} 2012, \apj, 750, 112

\bibitem[{{Liu} {et~al.}(2016){Liu}, {Yong}, {Asplund}, {Ram{\'{\i}}rez}, \&
  {Mel{\'e}ndez}}]{Liu2016}
{Liu}, F., {Yong}, D., {Asplund}, M., {Ram{\'{\i}}rez}, I., \& {Mel{\'e}ndez},
  J. 2016, \mnras, 457, 3934

\bibitem[{{Livingston} {et~al.}(2017){Livingston}, {Dai}, {Hirano}, {Gandolfi},
  {Nowak}, {Endl}, {Velasco}, {Fukui}, {Narita}, {Prieto-Arranz}, {Barragan},
  {Cusano}, {Albrecht}, {Cabrera}, {Cochran}, {Csizmadia}, {Deeg},
  {Eigm{\"u}ller}, {Erikson}, {Fridlund}, {Grziwa}, {Guenther}, {Hatzes},
  {Kawauchi}, {Korth}, {Nespral}, {Palle}, {P{\"a}tzold}, {Persson}, {Rauer},
  {Smith}, {Tamura}, {Tanaka}, {Van Eylen}, {Watanabe}, \&
  {Winn}}]{2017arXiv171007203L}
{Livingston}, J.~H., {Dai}, F., {Hirano}, T., {et~al.} 2017, ArXiv e-prints,
  arXiv:1710.07203

\bibitem[{{Lopez} {et~al.}(2012){Lopez}, {Fortney}, \&
  {Miller}}]{2012ApJ...761...59L}
{Lopez}, E.~D., {Fortney}, J.~J., \& {Miller}, N. 2012, \apj, 761, 59

\bibitem[{{Lundkvist} {et~al.}(2016){Lundkvist}, {Kjeldsen}, {Albrecht},
  {Davies}, {Basu}, {Huber}, {Justesen}, {Karoff}, {Silva Aguirre}, {van
  Eylen}, {Vang}, {Arentoft}, {Barclay}, {Bedding}, {Campante}, {Chaplin},
  {Christensen-Dalsgaard}, {Elsworth}, {Gilliland}, {Handberg}, {Hekker},
  {Kawaler}, {Lund}, {Metcalfe}, {Miglio}, {Rowe}, {Stello}, {Tingley}, \&
  {White}}]{2016NatCo...711201L}
{Lundkvist}, M.~S., {Kjeldsen}, H., {Albrecht}, S., {et~al.} 2016, Nature
  Communications, 7, 11201

\bibitem[{Mace {et~al.}(2016)Mace, Kim, Jaffe, Park, Lee, Kaplan, Yu, Yuk,
  Chun, Pak, Kim, Lee, Sneden, Afsar, Pavel, Lee, Oh, Jeong, Park, Kidder, Lee,
  Nguyen~Le, McLane, Gully-Santiago, Oh, Lee, Hwang, \& Park}]{Mace2016}
Mace, G., Kim, H., Jaffe, D.~T., {et~al.} 2016, in Society of Photo-Optical
  Instrumentation Engineers (SPIE) Conference Series, Vol. 9908, Proc. SPIE,
  99080C

\bibitem[{{Malavolta} {et~al.}(2016){Malavolta}, {Nascimbeni}, {Piotto},
  {Quinn}, {Borsato}, {Granata}, {Bonomo}, {Marzari}, {Bedin}, {Rainer},
  {Desidera}, {Lanza}, {Poretti}, {Sozzetti}, {White}, {Latham}, {Cunial},
  {Libralato}, {Nardiello}, {Boccato}, {Claudi}, {Cosentino}, {Covino},
  {Gratton}, {Maggio}, {Micela}, {Molinari}, {Pagano}, {Smareglia}, {Affer},
  {Andreuzzi}, {Aparicio}, {Benatti}, {Bignamini}, {Borsa}, {Damasso}, {Di
  Fabrizio}, {Harutyunyan}, {Esposito}, {Fiorenzano}, {Gandolfi}, {Giacobbe},
  {Gonz{\'a}lez Hern{\'a}ndez}, {Maldonado}, {Masiero}, {Molinaro}, {Pedani},
  \& {Scandariato}}]{2016A&A...588A.118M}
{Malavolta}, L., {Nascimbeni}, V., {Piotto}, G., {et~al.} 2016, \aap, 588, A118

\bibitem[{{Malo} {et~al.}(2013){Malo}, {Doyon}, {Lafreni{\`e}re}, {Artigau},
  {Gagn{\'e}}, {Baron}, \& {Riedel}}]{Malo2013}
{Malo}, L., {Doyon}, R., {Lafreni{\`e}re}, D., {et~al.} 2013, \apj, 762, 88

\bibitem[{{Mann} {et~al.}(2013{\natexlab{a}}){Mann}, {Brewer}, {Gaidos},
  {L{\'e}pine}, \& {Hilton}}]{Mann2013a}
{Mann}, A.~W., {Brewer}, J.~M., {Gaidos}, E., {L{\'e}pine}, S., \& {Hilton},
  E.~J. 2013{\natexlab{a}}, \aj, 145, 52

\bibitem[{{Mann} {et~al.}(2015){Mann}, {Feiden}, {Gaidos}, {Boyajian}, \& {von
  Braun}}]{Mann2015b}
{Mann}, A.~W., {Feiden}, G.~A., {Gaidos}, E., {Boyajian}, T., \& {von Braun},
  K. 2015, \apj, 804, 64

\bibitem[{{Mann} {et~al.}(2013{\natexlab{b}}){Mann}, {Gaidos}, {Kraus}, \&
  {Hilton}}]{Mann2013b}
{Mann}, A.~W., {Gaidos}, E., {Kraus}, A., \& {Hilton}, E.~J.
  2013{\natexlab{b}}, \apj, 770, 43

\bibitem[{{Mann} \& {von Braun}(2015)}]{Mann2015a}
{Mann}, A.~W., \& {von Braun}, K. 2015, \pasp, 127, 102

\bibitem[{{Mann} {et~al.}(2016){Mann}, {Gaidos}, {Mace}, {Johnson}, {Bowler},
  {LaCourse}, {Jacobs}, {Vanderburg}, {Kraus}, {Kaplan}, \&
  {Jaffe}}]{Mann2016a}
{Mann}, A.~W., {Gaidos}, E., {Mace}, G.~N., {et~al.} 2016, \apj, 818, 46

\bibitem[{{Mann} {et~al.}(2017{\natexlab{a}}){Mann}, {Dupuy}, {Muirhead},
  {Johnson}, {Liu}, {Ansdell}, {Dalba}, {Swift}, \& {Hadden}}]{Mann:2017ab}
{Mann}, A.~W., {Dupuy}, T., {Muirhead}, P.~S., {et~al.} 2017{\natexlab{a}},
  \aj, 153, 267

\bibitem[{{Mann} {et~al.}(2017{\natexlab{b}}){Mann}, {Gaidos}, {Vanderburg},
  {Rizzuto}, {Ansdell}, {Medina}, {Mace}, {Kraus}, \& {Sokal}}]{Mann:2017aa}
{Mann}, A.~W., {Gaidos}, E., {Vanderburg}, A., {et~al.} 2017{\natexlab{b}},
  \aj, 153, 64

\bibitem[{{Mazeh} {et~al.}(2013){Mazeh}, {Nachmani}, {Holczer}, {Fabrycky},
  {Ford}, {Sanchis-Ojeda}, {Sokol}, {Rowe}, {Zucker}, {Agol}, {Carter},
  {Lissauer}, {Quintana}, {Ragozzine}, {Steffen}, \&
  {Welsh}}]{2013ApJS..208...16M}
{Mazeh}, T., {Nachmani}, G., {Holczer}, T., {et~al.} 2013, \apjs, 208, 16

\bibitem[{{Meibom} {et~al.}(2011){Meibom}, {Barnes}, {Latham}, {Batalha},
  {Borucki}, {Koch}, {Basri}, {Walkowicz}, {Janes}, {Jenkins}, {Van Cleve},
  {Haas}, {Bryson}, {Dupree}, {Furesz}, {Szentgyorgyi}, {Buchhave}, {Clarke},
  {Twicken}, \& {Quintana}}]{2011ApJ...733L...9M}
{Meibom}, S., {Barnes}, S.~A., {Latham}, D.~W., {et~al.} 2011, \apjl, 733, L9

\bibitem[{{Meibom} {et~al.}(2013){Meibom}, {Torres}, {Fressin}, {Latham},
  {Rowe}, {Ciardi}, {Bryson}, {Rogers}, {Henze}, {Janes}, {Barnes}, {Marcy},
  {Isaacson}, {Fischer}, {Howell}, {Horch}, {Jenkins}, {Schuler}, \&
  {Crepp}}]{Meibom2013}
{Meibom}, S., {Torres}, G., {Fressin}, F., {et~al.} 2013, \nat, 499, 55

\bibitem[{{Monet} {et~al.}(2003){Monet}, {Levine}, {Canzian}, {Ables}, {Bird},
  {Dahn}, {Guetter}, {Harris}, {Henden}, {Leggett}, {Levison}, {Luginbuhl},
  {Martini}, {Monet}, {Munn}, {Pier}, {Rhodes}, {Riepe}, {Sell}, {Stone},
  {Vrba}, {Walker}, {Westerhout}, {Brucato}, {Reid}, {Schoening}, {Hartley},
  {Read}, \& {Tritton}}]{Monet:2003fj}
{Monet}, D.~G., {Levine}, S.~E., {Canzian}, B., {et~al.} 2003, \aj, 125, 984

\bibitem[{{Morton}(2015)}]{2015ascl.soft03011M}
{Morton}, T.~D. 2015, {VESPA: False positive probabilities calculator},
  Astrophysics Source Code Library, ascl:1503.011

\bibitem[{{Motalebi} {et~al.}(2015){Motalebi}, {Udry}, {Gillon}, {Lovis},
  {S{\'e}gransan}, {Buchhave}, {Demory}, {Malavolta}, {Dressing}, {Sasselov},
  {Rice}, {Charbonneau}, {Collier Cameron}, {Latham}, {Molinari}, {Pepe},
  {Affer}, {Bonomo}, {Cosentino}, {Dumusque}, {Figueira}, {Fiorenzano},
  {Gettel}, {Harutyunyan}, {Haywood}, {Johnson}, {Lopez}, {Lopez-Morales},
  {Mayor}, {Micela}, {Mortier}, {Nascimbeni}, {Philips}, {Piotto}, {Pollacco},
  {Queloz}, {Sozzetti}, {Vanderburg}, \& {Watson}}]{2015A&A...584A..72M}
{Motalebi}, F., {Udry}, S., {Gillon}, M., {et~al.} 2015, \aap, 584, A72

\bibitem[{{Mui{\~n}os} \& {Evans}(2014)}]{CMC15}
{Mui{\~n}os}, J.~L., \& {Evans}, D.~W. 2014, Astronomische Nachrichten, 335,
  367

\bibitem[{{Muirhead} {et~al.}(2015){Muirhead}, {Mann}, {Vanderburg}, {Morton},
  {Kraus}, {Ireland}, {Swift}, {Feiden}, {Gaidos}, \& {Gazak}}]{Muirhead2015}
{Muirhead}, P.~S., {Mann}, A.~W., {Vanderburg}, A., {et~al.} 2015, \apj, 801,
  18

\bibitem[{{Neves} {et~al.}(2012){Neves}, {Bonfils}, {Santos}, {Delfosse},
  {Forveille}, {Allard}, {Nat{\'a}rio}, {Fernandes}, \& {Udry}}]{Neves2012}
{Neves}, V., {Bonfils}, X., {Santos}, N.~C., {et~al.} 2012, \aap, 538, A25

\bibitem[{{Newton} {et~al.}(2015){Newton}, {Charbonneau}, {Irwin}, \&
  {Mann}}]{Newton2015A}
{Newton}, E.~R., {Charbonneau}, D., {Irwin}, J., \& {Mann}, A.~W. 2015, \apj,
  800, 85

\bibitem[{{Obermeier} {et~al.}(2016){Obermeier}, {Henning}, {Schlieder},
  {Crossfield}, {Petigura}, {Howard}, {Sinukoff}, {Isaacson}, {Ciardi},
  {David}, {Hillenbrand}, {Beichman}, {Howell}, {Horch}, {Everett}, {Hirsch},
  {Teske}, {Christiansen}, {L{\'e}pine}, {Aller}, {Liu}, {Saglia},
  {Livingston}, \& {Kluge}}]{Obermeier:2016aa}
{Obermeier}, C., {Henning}, T., {Schlieder}, J.~E., {et~al.} 2016, \aj, 152,
  223

\bibitem[{{Oshagh} {et~al.}(2017){Oshagh}, {Santos}, {Figueira}, {Barros},
  {Donati}, {Adibekyan}, {Faria}, {Watson}, {Cegla}, {Dumusque}, {H{\'e}brard},
  {Demangeon}, {Dreizler}, {Boisse}, {Deleuil}, {Bonfils}, {Pepe}, \&
  {Udry}}]{Oshagh:2017aa}
{Oshagh}, M., {Santos}, N.~C., {Figueira}, P., {et~al.} 2017, \aap, 606, A107

\bibitem[{{Palmer} {et~al.}(2014){Palmer}, {Arenou}, {Luri}, \&
  {Masana}}]{2014A&A...564A..49P}
{Palmer}, M., {Arenou}, F., {Luri}, X., \& {Masana}, E. 2014, \aap, 564, A49

\bibitem[{{Park} {et~al.}(2014){Park}, {Jaffe}, {Yuk}, {Chun}, {Pak}, {Kim},
  {Pavel}, {Lee}, {Oh}, {Jeong}, {Sim}, {Lee}, {Nguyen Le}, {Strubhar},
  {Gully-Santiago}, {Oh}, {Cha}, {Moon}, {Park}, {Brooks}, {Ko}, {Han}, {Nah},
  {Hill}, {Lee}, {Barnes}, {Yu}, {Kaplan}, {Mace}, {Kim}, {Lee}, {Hwang}, \&
  {Park}}]{Park2014}
{Park}, C., {Jaffe}, D.~T., {Yuk}, I.-S., {et~al.} 2014, in Society of
  Photo-Optical Instrumentation Engineers (SPIE) Conference Series, Vol. 9147,
  Society of Photo-Optical Instrumentation Engineers (SPIE) Conference Series,
  1

\bibitem[{{Parviainen} \& {Aigrain}(2015)}]{2015MNRAS.453.3821P}
{Parviainen}, H., \& {Aigrain}, S. 2015, \mnras, 453, 3821

\bibitem[{{Paulson} {et~al.}(2003){Paulson}, {Sneden}, \&
  {Cochran}}]{2003AJ....125.3185P}
{Paulson}, D.~B., {Sneden}, C., \& {Cochran}, W.~D. 2003, \aj, 125, 3185

\bibitem[{{Pecaut} {et~al.}(2012){Pecaut}, {Mamajek}, \&
  {Bubar}}]{2012ApJ...746..154P}
{Pecaut}, M.~J., {Mamajek}, E.~E., \& {Bubar}, E.~J. 2012, \apj, 746, 154

\bibitem[{{Pepe} {et~al.}(2013){Pepe}, {Cameron}, {Latham}, {Molinari}, {Udry},
  {Bonomo}, {Buchhave}, {Charbonneau}, {Cosentino}, {Dressing}, {Dumusque},
  {Figueira}, {Fiorenzano}, {Gettel}, {Harutyunyan}, {Haywood}, {Horne},
  {Lopez-Morales}, {Lovis}, {Malavolta}, {Mayor}, {Micela}, {Motalebi},
  {Nascimbeni}, {Phillips}, {Piotto}, {Pollacco}, {Queloz}, {Rice}, {Sasselov},
  {S{\'e}gransan}, {Sozzetti}, {Szentgyorgyi}, \&
  {Watson}}]{2013Natur.503..377P}
{Pepe}, F., {Cameron}, A.~C., {Latham}, D.~W., {et~al.} 2013, \nat, 503, 377

\bibitem[{{Pepper} {et~al.}(2017){Pepper}, {Gillen}, {Parviainen},
  {Hillenbrand}, {Cody}, {Aigrain}, {Stauffer}, {Vrba}, {David}, {Lillo-Box},
  {Stassun}, {Conroy}, {Pope}, \& {Barrado}}]{2017AJ....153..177P}
{Pepper}, J., {Gillen}, E., {Parviainen}, H., {et~al.} 2017, \aj, 153, 177

\bibitem[{{Quinn} {et~al.}(2014){Quinn}, {White}, {Latham}, {Buchhave},
  {Torres}, {Stefanik}, {Berlind}, {Bieryla}, {Calkins}, {Esquerdo}, {F{\H
  u}r{\'e}sz}, {Geary}, \& {Szentgyorgyi}}]{Quinn2014}
{Quinn}, S.~N., {White}, R.~J., {Latham}, D.~W., {et~al.} 2014, \apj, 787, 27

\bibitem[{{Rayner} {et~al.}(2009){Rayner}, {Cushing}, \& {Vacca}}]{Rayner2009}
{Rayner}, J.~T., {Cushing}, M.~C., \& {Vacca}, W.~D. 2009, \apjs, 185, 289

\bibitem[{{Rizzuto} {et~al.}(2015){Rizzuto}, {Ireland}, \&
  {Kraus}}]{rizzuto2015}
{Rizzuto}, A.~C., {Ireland}, M.~J., \& {Kraus}, A.~L. 2015, \mnras, 448, 2737

\bibitem[{{Rizzuto} {et~al.}(2011){Rizzuto}, {Ireland}, \&
  {Robertson}}]{rizzuto2011}
{Rizzuto}, A.~C., {Ireland}, M.~J., \& {Robertson}, J.~G. 2011, \mnras, 416,
  3108

\bibitem[{{Rizzuto} {et~al.}(2017){Rizzuto}, {Mann}, {Vanderburg}, \&
  {Kraus}}]{2017arXiv170909670R}
{Rizzuto}, A.~C., {Mann}, A.~W., {Vanderburg}, A., \& {Kraus}, A.~L. 2017,
  ArXiv e-prints, arXiv:1709.09670

\bibitem[{{Roeser} {et~al.}(2010){Roeser}, {Demleitner}, \&
  {Schilbach}}]{2010AJ....139.2440R}
{Roeser}, S., {Demleitner}, M., \& {Schilbach}, E. 2010, \aj, 139, 2440

\bibitem[{{Rogers}(2015)}]{2015ApJ...801...41R}
{Rogers}, L.~A. 2015, \apj, 801, 41

\bibitem[{{R{\"o}ser} {et~al.}(2011){R{\"o}ser}, {Schilbach}, {Piskunov},
  {Kharchenko}, \& {Scholz}}]{Roser2011}
{R{\"o}ser}, S., {Schilbach}, E., {Piskunov}, A.~E., {Kharchenko}, N.~V., \&
  {Scholz}, R.-D. 2011, \aap, 531, A92

\bibitem[{{Seager} \& {Mall{\'e}n-Ornelas}(2003)}]{Seager:2003lr}
{Seager}, S., \& {Mall{\'e}n-Ornelas}, G. 2003, \apj, 585, 1038

\bibitem[{{Sinukoff} {et~al.}(2017){Sinukoff}, {Howard}, {Petigura}, {Fulton},
  {Crossfield}, {Isaacson}, {Gonzales}, {Crepp}, {Brewer}, {Hirsch}, {Weiss},
  {Ciardi}, {Schlieder}, {Benneke}, {Christiansen}, {Dressing}, {Hansen},
  {Knutson}, {Kosiarek}, {Livingston}, {Greene}, {Rogers}, \&
  {L{\'e}pine}}]{2017AJ....153..271S}
{Sinukoff}, E., {Howard}, A.~W., {Petigura}, E.~A., {et~al.} 2017, \aj, 153,
  271

\bibitem[{{Skrutskie} {et~al.}(2006){Skrutskie}, {Cutri}, {Stiening},
  {Weinberg}, {Schneider}, {Carpenter}, {Beichman}, {Capps}, {Chester},
  {Elias}, {Huchra}, {Liebert}, {Lonsdale}, {Monet}, {Price}, {Seitzer},
  {Jarrett}, {Kirkpatrick}, {Gizis}, {Howard}, {Evans}, {Fowler}, {Fullmer},
  {Hurt}, {Light}, {Kopan}, {Marsh}, {McCallon}, {Tam}, {Van Dyk}, \&
  {Wheelock}}]{Skrutskie2006}
{Skrutskie}, M.~F., {Cutri}, R.~M., {Stiening}, R., {et~al.} 2006, \aj, 131,
  1163

\bibitem[{{Stumpe} {et~al.}(2012){Stumpe}, {Smith}, {Van Cleve}, {Twicken},
  {Barclay}, {Fanelli}, {Girouard}, {Jenkins}, {Kolodziejczak}, {McCauliff}, \&
  {Morris}}]{Stumpe2012}
{Stumpe}, M.~C., {Smith}, J.~C., {Van Cleve}, J.~E., {et~al.} 2012, \pasp, 124,
  985

\bibitem[{{Swift} {et~al.}(2015){Swift}, {Montet}, {Vanderburg}, {Morton},
  {Muirhead}, \& {Johnson}}]{Swift2015}
{Swift}, J.~J., {Montet}, B.~T., {Vanderburg}, A., {et~al.} 2015, \apjs, 218,
  26

\bibitem[{{Twicken} {et~al.}(2010){Twicken}, {Chandrasekaran}, {Jenkins},
  {Gunter}, {Girouard}, \& {Klaus}}]{2010SPIE.7740E..1UT}
{Twicken}, J.~D., {Chandrasekaran}, H., {Jenkins}, J.~M., {et~al.} 2010, in
  \procspie, Vol. 7740, Software and Cyberinfrastructure for Astronomy, 77401U

\bibitem[{{Vacca} {et~al.}(2003){Vacca}, {Cushing}, \& {Rayner}}]{Vacca2003}
{Vacca}, W.~D., {Cushing}, M.~C., \& {Rayner}, J.~T. 2003, \pasp, 115, 389

\bibitem[{{Van Eylen} \& {Albrecht}(2015)}]{Van-Eylen2015}
{Van Eylen}, V., \& {Albrecht}, S. 2015, \apj, 808, 126

\bibitem[{{van Leeuwen}(2009)}]{van-Leeuwen2009}
{van Leeuwen}, F. 2009, \aap, 497, 209

\bibitem[{{Vanderburg} \& {Johnson}(2014)}]{Vanderburg2014}
{Vanderburg}, A., \& {Johnson}, J.~A. 2014, \pasp, 126, 948

\bibitem[{{Vanderburg} {et~al.}(2015){Vanderburg}, {Johnson}, {Rappaport},
  {Bieryla}, {Irwin}, {Lewis}, {Kipping}, {Brown}, {Dufour}, {Ciardi}, {Angus},
  {Schaefer}, {Latham}, {Charbonneau}, {Beichman}, {Eastman}, {McCrady},
  {Wittenmyer}, \& {Wright}}]{2015Natur.526..546V}
{Vanderburg}, A., {Johnson}, J.~A., {Rappaport}, S., {et~al.} 2015, \nat, 526,
  546

\bibitem[{{Vanderburg} {et~al.}(2016){Vanderburg}, {Latham}, {Buchhave},
  {Bieryla}, {Berlind}, {Calkins}, {Esquerdo}, {Welsh}, \& {Johnson}}]{v16}
{Vanderburg}, A., {Latham}, D.~W., {Buchhave}, L.~A., {et~al.} 2016, \apjs,
  222, 14

\bibitem[{Vanderplas(2015)}]{jake_vanderplas_2015_14475}
Vanderplas, J. 2015, {supersmoother: Efficient Python Implementation of
  Friedman's SuperSmoother}, doi:10.5281/zenodo.14475

\bibitem[{{Veras} {et~al.}(2011){Veras}, {Ford}, \& {Payne}}]{Veras:2011nx}
{Veras}, D., {Ford}, E.~B., \& {Payne}, M.~J. 2011, \apj, 727, 74

\bibitem[{{Weiss} \& {Marcy}(2014)}]{2014ApJ...783L...6W}
{Weiss}, L.~M., \& {Marcy}, G.~W. 2014, \apjl, 783, L6

\bibitem[{{Wright} {et~al.}(2010){Wright}, {Eisenhardt}, {Mainzer}, {Ressler},
  {Cutri}, {Jarrett}, {Kirkpatrick}, {Padgett}, {McMillan}, {Skrutskie},
  {Stanford}, {Cohen}, {Walker}, {Mather}, {Leisawitz}, {Gautier}, {McLean},
  {Benford}, {Lonsdale}, {Blain}, {Mendez}, {Irace}, {Duval}, {Liu}, {Royer},
  {Heinrichsen}, {Howard}, {Shannon}, {Kendall}, {Walsh}, {Larsen}, {Cardon},
  {Schick}, {Schwalm}, {Abid}, {Fabinsky}, {Naes}, \& {Tsai}}]{Wright2010}
{Wright}, E.~L., {Eisenhardt}, P.~R.~M., {Mainzer}, A.~K., {et~al.} 2010, \aj,
  140, 1868

\bibitem[{{Yuk} {et~al.}(2010){Yuk}, {Jaffe}, {Barnes}, {Chun}, {Park}, {Lee},
  {Lee}, {Wang}, {Park}, {Pak}, {Strubhar}, {Deen}, {Oh}, {Seo}, {Pyo}, {Park},
  {Lacy}, {Goertz}, {Rand}, \& {Gully-Santiago}}]{2010SPIE.7735E..1MY}
{Yuk}, I.-S., {Jaffe}, D.~T., {Barnes}, S., {et~al.} 2010, in Society of
  Photo-Optical Instrumentation Engineers (SPIE) Conference Series, Vol. 7735,
  Society of Photo-Optical Instrumentation Engineers (SPIE) Conference Series

\bibitem[{{Zacharias} {et~al.}(2017){Zacharias}, {Finch}, \&
  {Frouard}}]{2017AJ....153..166Z}
{Zacharias}, N., {Finch}, C., \& {Frouard}, J. 2017, \aj, 153, 166

\bibitem[{{Zacharias} {et~al.}(2015){Zacharias}, {Finch}, {Subasavage},
  {Bredthauer}, {Crockett}, {Divittorio}, {Ferguson}, {Harris}, {Harris},
  {Henden}, {Kilian}, {Munn}, {Rafferty}, {Rhodes}, {Schultheiss}, {Tilleman},
  \& {Wieder}}]{URAT1}
{Zacharias}, N., {Finch}, C., {Subasavage}, J., {et~al.} 2015, \aj, 150, 101

\end{thebibliography}

\clearpage
\floattable
\begin{deluxetable}{l c l }
\tabletypesize{\scriptsize}
\tablecaption{Parameters of \tar\ \label{tab:stellar}}
\tablewidth{0pt}
\tablehead{
\colhead{Parameter} & \colhead{Value} & \colhead{Source}
}
\startdata
\multicolumn{3}{c}{\hspace{1cm}Astrometry} \\
$\alpha$ R.A. (hh:mm:ss) & 04:29:38.99 & EPIC\\
$\delta$ Dec. (dd:mm:ss) & +22:52:57.8 & EPIC\\
$\mu_{\alpha}$ (mas~yr$^{-1}$) & $+83.0\pm0.9$  & HSOY \\
$\mu{\delta}$ (mas~yr$^{-1}$) & $-35.7\pm0.9$ & HSOY \\
\hline
\multicolumn {3}{c}{\hspace{1cm}Photometry} \\
$B$ (mag) & 12.48 $\pm$ 0.08 & APASS\tablenotemark{a} \\
$V$ (mag) & 11.20 $\pm$ 0.08 & APASS\tablenotemark{a} \\
$g$ (mag) & 11.93 $\pm$ 0.08 & APASS\tablenotemark{a} \\			
$r$ (mag) & 10.74 $\pm$ 0.08 & APASS\tablenotemark{a} \\			
$i$ (mag) & 10.25 $\pm$ 0.08 & APASS\tablenotemark{a} \\
$G_{Gaia}$ (mag) & 10.747 $\pm$ 0.001 & GaiaDR1 \\
$r$ (mag) & 10.823 $\pm$ 0.007 & CMC15 \\
$J$ (mag) & 9.096 $\pm$ 0.022 & 2MASS\\
$H$ (mag) & 8.496 $\pm$ 0.020 & 2MASS\\
$K_s$ (mag) & 8.368 $\pm$ 0.019 & 2MASS\\ 					
$W1$ (mag) & 8.263 $\pm$ 0.023 & $WISE$ \\
$W2$ (mag) & 8.496 $\pm$ 0.020 & $WISE$\\
$W3$ (mag) & 8.368 $\pm$ 0.019 & $WISE$ \\
\hline
\multicolumn{3}{c}{\hspace{1cm}Kinematics/Position} \\
Barycentric RV (\kms) & 39.76 $\pm$ 0.10 & This paper \\
$U$ (\kms) & $-43.5\pm  0.8$ & This paper\\
$V$ (\kms) & $-21.2\pm  3.6$ & This paper \\
$W$ (\kms) & $-0.3\pm  1.7$ & This paper \\
$X$ (pc) & $-59.9\pm  9.5$ & This paper \\
$Y$ (pc) & $  +5.5\pm  0.9$ & This paper \\
$Z$ (pc) & $-18.8\pm  3.0$ & This paper \\
Kinematic Distance (pc) &   $59.4 \pm 2.8$ & This paper\\
Photometric Distance (pc) & $63 \pm 10$ & This paper\\
\hline
\multicolumn{3}{c}{\hspace{1cm}Physical Properties} \\
Rotation period (days) & 15.0  $\pm$ 1.0 & This paper \\
\vsini\ (\kms) & $3.0\pm0.5$  & This paper \\
Spectral Type & K5.5 $\pm$ 0.5 & This paper \\
$[$Fe/H$]$ (dex) & 0.15 $\pm$ 0.03 & Hyades Value\\
\teff\ (K) &  4499$\pm$50 & This paper\\
$M_*$ ($M_\odot$) &  0.74$\pm$0.02  & This paper \\
$R_*$ ($R_\odot$) &  0.66$\pm$0.02 & This paper \\
$L_*$ ($L_\odot$) & 0.163$\pm$0.016 & This paper \\
$\rho_*$ ($\rho_\odot$) &  $2.50^{+0.13}_{-0.12}$ & This paper\\
log~g (dex) &  $4.66\pm0.02$ & This paper\\
\enddata
\tablenotetext{a}{Reported APASS magnitude errors are from a single-epoch only; so we adopted 0.08\,mag errors based on a comparison to similar catalogs. Gaia G magnitude error taken from flux measurement standard deviation.}
\end{deluxetable}

\floattable
\begin{deluxetable}{l c c c l }
\tabletypesize{\scriptsize}
\tablecaption{Radial Velocities of \tar\ \label{tab:rv}}
\tablewidth{0pt}
\tablehead{
\colhead{Year} & \colhead{HJD-2400000} & \colhead{RV (km/s)} & \colhead{$\sigma_\mathrm{RV}$ (km/s)} & \colhead{Source}
}
\startdata
1983.87 &  45651.92610 & 39.51 & 0.50 & DSP\tablenotemark{a}\\
1984.07 &  45724.63660 & 39.22 & 0.53 & DSP\tablenotemark{a}\\
1989.16 &  47585.62560 & 39.61 & 0.44 & DSP\tablenotemark{a}\\
1993.93 &  49326.74090 & 40.36 & 0.40 & DSP\tablenotemark{a}\\
2004.07 &  53029.56670 & 39.90 & 0.47 & DSP\tablenotemark{a}\\
2017.67 &  57998.95220 & 39.73 & 0.10 & TRES\\
2017.67 &  57999.00020 & 39.10 & 0.20 & IGRINS\\
2017.68 &  58000.95080 & 39.76 & 0.10 & TRES\\
\enddata
\tablenotetext{a}{CfA Digital Speedometers}
\end{deluxetable}

\floattable
\begin{deluxetable*}{l l l l l  }
\tabletypesize{\scriptsize}
\tablecaption{Transit-Fit Parameters \label{tab:planet}}
\tablewidth{0pt}
\tablehead{
\colhead{Parameter}&\colhead{Planet b}&\colhead{Planet c}&\colhead{Planet d}
}
\startdata
Period (days)  & $7.975292^{+0.000833}_{-0.000770}$ & $17.307137^{+0.000252}_{-0.000284}$ & $25.575065^{+0.002418}_{-0.002357}$\\
$R_P/R_*$  & $0.0137^{+0.0006}_{-0.0004}$ & $0.0401^{+0.0012}_{-0.0006}$ & $0.0201^{+0.0014}_{-0.0009}$\\
T$_0$ (BJD-2400000)  & $57817.75631^{+0.00469}_{-0.00489}$ & $57812.71770^{+0.00100}_{-0.00077}$ & $57780.81164^{+0.00643}_{-0.00660}$\\
Impact Parameter & $0.27^{+0.28}_{-0.19}$  & $0.28^{+0.21}_{-0.18}$ & $0.53^{+0.19}_{-0.33}$\\
Duration\tablenotemark{a} (hours) & $2.54^{+0.24}_{-0.49}$  & $3.45^{+0.60}_{-0.69}$ & $3.12^{+1.30}_{-0.54}$\\
Inclination\tablenotemark{a} (degrees) & $89.3^{+0.5}_{-0.7}$ & $89.6^{+0.3}_{-0.3}$ & $89.4^{+0.4}_{-0.2}$\\
$a/R_*$\tablenotemark{a} & $23.1^{+2.4}_{-1.2}$  & $39.0^{+7.2}_{-1.5}$  & $50.7^{+5.2}_{-4.5}$ \\
Eccentricity  & $0.10^{+  0.19}_{-0.07}$  & $0.13^{+  0.27}_{-0.11}$  & $0.14^{+  0.13}_{-0.09}$ \\
$\omega$ (degrees)  & $12^{+145}_{-70}$  & $24^{+141}_{-74}$  & $2^{+ 91}_{-148}$ \\
$R_P$\tablenotemark{c} (R$_\earth$): 	&	$0.99^{+0.06}_{-0.04}$ & $2.91^{+0.11}_{-0.10}$ & $1.45^{+0.11}_{-0.08}$\\
\teq\tablenotemark{b} (K): & $553^{+17}_{-27}$ & $425^{+10}_{-33}$ & $373^{+18}_{-17}$&\\
\hline
\multicolumn{4}{c}{Global Parameters}\\
\hline
$\rho_*$ ($\rho_\odot$)  &  & $2.57^{+0.15}_{-0.16}$ &      \\
$u_1$  &  & $0.65^{+0.05}_{-0.05}$  &      \\
$u_2$  &  & $0.10^{+0.06}_{-0.06}$  &      \\
\enddata
\tablenotetext{a}{ Inclination, $a/R_*$, eccentricity, and $\omega$, are not fit as part of the MCMC, but are instead derived from the other fit parameters (see Section~\ref{sec:transit}). We report them here for convenience. }
\tablenotetext{b}{$R_P$ and \teq\ use the $R_*$ or \teff\ value from Section~\ref{sec:params}. \teq\ calculation assumes an albedo of exactly 0.3 for simplicity.}
\end{deluxetable*}

\end{document}